\documentclass[preprint, 10pt]{elsarticle}

\newif\ifInputs
\Inputstrue

\usepackage{palatino}
\usepackage[usenames]{color}
\usepackage{epsfig}
\usepackage[fleqn,reqno]{amsmath}
\usepackage{amsfonts,amsthm,bm}
\usepackage{amssymb}
\usepackage{mathrsfs}
\usepackage{stmaryrd}
\usepackage{subfigure}
\usepackage[titletoc]{appendix}
\usepackage{tabularx,booktabs}
\usepackage{graphics,caption}
\usepackage{fancybox}
\usepackage[top=1.2in,bottom=1.2in,left=1in, right=1in]{geometry}
\usepackage{array}
\usepackage{multirow}
\usepackage{wrapfig}
\usepackage{algorithmic,algorithm}
\usepackage{paralist}
\usepackage{ifthen}
\usepackage{enumitem}
\usepackage{tikz,pgfplots,filecontents}
\usepackage{float}
\usepackage{mdframed}
\usepackage{todonotes}
\usepackage{xcolor,colortbl}
\usepackage{soul}

\graphicspath{{./figs/}}

\usepackage[pagebackref=false,bookmarks=false]{hyperref} 

\definecolor{dblue}{rgb}{0.03,0.3,0.62}
\definecolor{dorange}{rgb}{1,0.55,0}

\setstcolor{red}

\hypersetup{
  bookmarksnumbered=true,
  bookmarksopen=false,
  hypertexnames=false,     
  breaklinks=true,          
  unicode=false,
  pdffitwindow=true,        
  pdfnewwindow=true,        
  colorlinks=true,         
  linkcolor=dblue,
  anchorcolor=red,
  citecolor=dorange,
  filecolor=magenta,
  urlcolor=dblue,
  pdfstartview = FitH,
  pdfkeywords = {},
  pdfcreator = {LaTeX with hyperref package}
}

\definecolor{sblue}{cmyk}{0.98,0.13,0,0.43} 
\definecolor{sblue}{cmyk}{0.98,0.13,0,0.43} 

\newcommand{\editS}[1]{{#1}}
\newcommand{\editnewS}[1]{{#1}}

\newcommand{\bigO}{\mathcal{O}}
\newcommand{\Div}{\mathrm{div}}

\newcommand{\figref}[1]{Figure~\ref{#1}}
\newcommand{\tabref}[1]{Table~\ref{#1}}
\newcommand{\mcaption}[2]{\caption{\small \em #1}\label{#2}}
\newcommand{\secref}[1]{Section \ref{#1}}
\newcommand{\algref}[1]{Algorithm \ref{#1}}
\newcommand{\appref}[1]{Appendix \ref{#1}}

\newcommand{\UU}{{\boldsymbol{U}}}
\newcommand{\Lap}{\rotatebox[origin=c]{180}{$\nabla$}}
\newcommand{\xx}{{\mathbf{x}}}

\newcommand{\yy}{{\mathbf{y}}}

\newcommand{\uu}{{\mathbf{u}}}

\newcommand{\rr}{{\mathbf{r}}}
\newcommand{\nn}{{\mathbf{n}}}

\newcommand{\zzeta}{{\boldsymbol\zeta}}

\newcommand{\Grad}{{\nabla}}

\newcolumntype{C}{>{\centering\arraybackslash} m{2.5cm}}

\usepackage{lineno}

\begin{document}

\title{Optimal design of deterministic lateral displacement device for viscosity contrast based cell sorting}

\author[utMe]{G\"{o}kberk Kabacao\u{g}lu} \ead{gokberk@ices.utexas.edu}
\author[utMe,ut]{George Biros}\ead{gbiros@acm.org}
\address[utMe]{Department of Mechanical Engineering, \\The University 
of Texas at Austin, Austin, TX, 78712, United States}
\address[ut]{Institute for Computational Engineering and Sciences,\\
The University of Texas at Austin, Austin, TX, 78712, United States}

\begin{abstract} 
We solve a design optimization problem for deterministic lateral displacement (DLD) device to sort same-size biological cells by their deformability, in particular to sort red blood cells (RBCs) by their viscosity contrast between the fluid in the interior and the exterior of the cells. A DLD device optimized for efficient cell sorting enables rapid medical diagnoses of several diseases such as malaria since infected cells are stiffer than their healthy counterparts. The device consists of pillar arrays in which pillar rows are tilted and hence are not orthogonal to the columns. This arrangement leads cells to have different final vertical displacements depending on their deformability, therefore, it vertically separates the cells. Pillar cross section, tilt angle of the pillar rows and center-to-center distances between pillars are free design parameters. For a given pair of viscosity contrast values of the cells we seek optimal DLD designs by fixing the tilt angle and the center-to-center distances. So the only design parameter is the pillar cross section. \editS{We propose an objective function such that a design minimizing it delivers designs providing efficient cell sorting.} The objective function is evaluated by simulating the cell flows through a device using our 2D model (Kabacaoglu et al. [\textit{Journal of Computational Physics}, \textbf{357}:43-77, 2018]). We solve the optimization problem using \editS{a stochastic optimization algorithm}. Since the algorithm converges in $\bigO(1000)$ iterations and our high-fidelity DLD model is expensive to evaluate the objective function, we propose a low-fidelity DLD model to enable fast solution of the problem. Finally, we present several scenarios where solving the optimization problem finds designs that can separate cells with similar viscosity contrast values. These designs have cross sections that have features similar to a triangle. To the best of our knowledge, this is the first study which poses designing a DLD device as a constrained optimization problem so as to discover optimal designs systematically.

\end{abstract}


\maketitle


\section{Introduction\label{s:intro}}
\begin{figure}[!htb]
\begin{minipage}{\textwidth}
\setcounter{subfigure}{0}
\centering
\renewcommand*{\thesubfigure}{(a)} 
      \hspace{0cm}\subfigure[Conventional DLD design cannot sort the cells since both cells zig-zag.]{\scalebox{0.75}{{\includegraphics{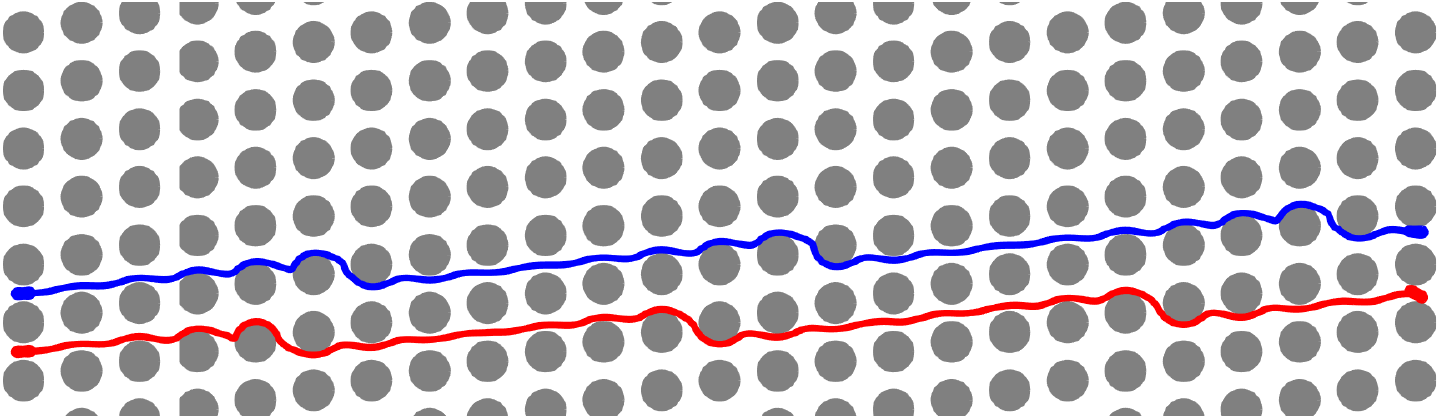}}}
      \label{f:VCs5a50CircMotiv}} 
\end{minipage}
\begin{minipage}{\textwidth}
\setcounter{subfigure}{0}      
\centering
\renewcommand*{\thesubfigure}{(b)} 
      \hspace{0cm}\subfigure[A design with a triangular pillar cross section and narrower gaps can sort the cells.]{\scalebox{0.75}{{\includegraphics{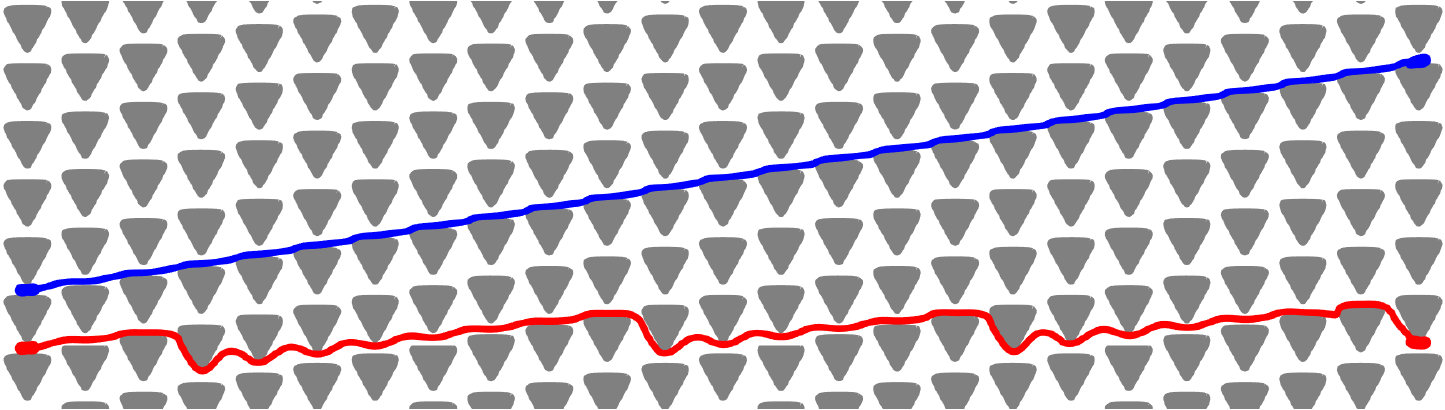}}}
      \label{f:VCs5a50TriMotiv}}
\end{minipage}                    
\mcaption{We demonstrate the top views of two DLD designs for deformability-based cell sorting. Flow is from left to right. Vertically aligned pillars (in gray) form columns and the pillar rows are tilted with respect to the flow axis. The red cell is stiffer than the blue cell. The conventional design in~\figref{f:VCs5a50CircMotiv} cannot sort the cells because it leads both to {\bf "zig-zag''}. Solving a design optimization problem guides designing the device in~\figref{f:VCs5a50TriMotiv}. The proposed design has the same tilt angle of the pillar rows as in~\figref{f:VCs5a50CircMotiv} but a triangular pillar cross section and narrower gaps. The new design enables cell sorting since it leads the soft cell to move along the tilted pillar rows (i.e., in the {\bf "displacement''} mode) and the stiff cell to zig-zag.  
}{f:motivation}
\end{figure}

Sorting biological cells by their sizes and/or deformability using lab-on-a-chip technology is a key step in rapid medical diagnoses and tests. For example, deformability-based sorting enables identifying malaria-infected red blood cells (RBCs) from whole blood since the diseased cells are much stiffer than their healthy counterparts~\cite{suresh-seufferlein-e05}. Deterministic lateral displacement (DLD) is a microfluidic sorting technique introduced recently by~\citet{huang-sturm-e04}. Due to its low cost and easy operation it has been used in many applications such as separation of cancer cells from blood~\cite{loutherback-sturm-e12,karabacak-toner-e14}, fractionation of blood components (white and red blood cells, platelets)~\cite{davis-austin-e06,inglis-sturm-e08,beech-tegenfeldt-e12} and separation of parasites from blood~\cite{holm-tegenfeldt-e11,holm-tegenfeldt-e16}. A DLD device consists of arrays of pillars (see~\figref{f:motivation} for the top view of devices with a cylindrical and a triangular pillar cross sections). The pillar grid forms a lattice but the lattice vectors are not orthogonal. That is, the pillars are vertically aligned but the pillar rows are tilted at an angle with the flow axis. Particle sorting takes place as a result of the hydrodynamic interactions between the particles and the pillars as follows. The pillar arrangement divides the flow in a vertical gap into a number of streams of equal mass flux. The widths of the streams depend on the pillar cross section, the sizes of the gaps between the pillars and the tilt angle of the pillar rows. Let us consider the pillar arrangement in~\figref{f:motivation}. Here, the flow is from left to right. The stream adjacent to each pillar (i.e., adjacent stream) flows downwards through the tilted horizontal gap while the other streams move along the tilted pillar rows. The particles that are trapped in the adjacent streams also flow downwards and those that can stay out of the these streams move along the tilted pillar rows. The former transport mode is called {\bf "zig-zag''} and the latter is called {\bf "displacement''}. For example, both cells in~\figref{f:VCs5a50CircMotiv} zig-zag while the blue one in~\figref{f:VCs5a50TriMotiv} displaces. The adjacent streams are continuously replaced by the other streams in the gap during the process. After several particle-pillar interactions, the displacing particle is vertically separated from the zig-zagging one since the latter has almost zero net vertical displacement. The analytical DLD theory for size-based sorting of rigid spherical particles has been developed~\cite{davis-austin-e06}. Yet, deformability-based sorting of biological cells is more complex phenomenon than that since it depends not only on the device geometry and the cells' size and orientation but also on the cells' rich dynamics such as migration and tumbling. 

DLD devices for sorting RBCs are called shallow if the pillars are shorter than an RBC's diameter (which is about $8\mu m$) and deep if they are much taller than the diameter. The deep devices can provide higher-throughput than the shallow ones. In this study we consider deep DLD devices and focus on flow regimes and particle properties that resemble RBCs. In such devices the cells' effective size is their thickness (which is about $3\mu m$) and they are free to show rich dynamics depending on their deformability. Their transport modes depend on how much they can migrate vertically or whether they can tumble. Vertical migration leads a cell to displace by keeping it away from the adjacent stream~\cite{henry-gompper-e16,kabacaoglu-biros17}. Tumbling increases the cell's effective size but reduces the migration. If the combined effect of these on the tumbling cell causes it to stay away from the adjacent stream, it displaces. There are a few studies on deformability-based sorting of RBCs in deep DLD devices~\cite{zhang-fedosov-e15,henry-gompper-e16}, including ours~\cite{kabacaoglu-biros17}. Both ours and~\cite{henry-gompper-e16} considered devices with circular pillar cross sections and aimed to explain how cell sorting takes place. \citet{zhang-fedosov-e15} investigated cell dynamics for triangular, square and diamond pillar cross sections. So, these recent studies are concerned with discovering cell dynamics in conventional DLD devices. Here we want to discover different DLD designs for efficient cell sorting. One of the difficulties is sorting cells with similar mechanical properties. That might either be impossible due to low sorting resolution or require long devices to induce sufficient vertical separation, which increases the process time. So, we need to design a device that is short but still capable of sorting such cells. Exhaustive search for that purpose is not practical because one needs to perform experiments or simulations to determine the cells' dynamics in every device. Also such computations are very expensive. How can we systematically design DLD devices for particular objectives and constraints? This is the main question we aim to address in this article.

\subsection{Methodology}
In~\cite{kabacaoglu-biros17}, we presented a numerical scheme for 2D simulations of flows of RBCs in DLD devices. The scheme has no free parameters other than the time-step size and spatial discretization. It can reproduce numerical and experimental results and phase diagrams in both two and three dimensions. In this study, we use the same numerical scheme. Let us briefly revisit this model. An RBC is modeled as an inextensible \editS{vesicle} with a biconcave shape that resists deformation due to bending and tension~\cite{keller-skalak82,misbah12}. It is impermeable to flow. The fluid in the interior and exterior of the cell is Newtonian. An RBC's deformability depends on its bending rigidity, the imposed shear rate, the viscosities of the fluid in the interior and the exterior of the cell. The deformability can be characterized by two dimensionless numbers: the capillary number $C_a$ and the viscosity contrast $\nu$. In this paper, we adjust $\nu$ only. \editnewS{We use a standard quasi-static Stokes approximation scheme to model the flow~\cite{kraus-lipowsky-e96,danker-misbah-e09,zhao-freund-e10}, and formulate the problem as a set of integro-differential equations~\cite{rallison-acrivos78,pozrikidis92,kraus-lipowsky-e96,shravan-biros-e09,biben-misbah-e11}.} 

\editS{Design parameters of a DLD device are pillar cross section (i.e., top view of pillars), tilt angle of pillar rows and center-to-center distances. These define a unique device.} We fix the tilt angle and the center-to-center distances. So, the only design parameter is pillar cross section which we parameterize with uniform $5^{th}$ order B-splines (See~\appref{a:bSplineCoeffs}). The objective function \editS{for the optimization problem} assesses whether a design provides efficient cell sorting but quantifying this statement is not obvious. We discuss the choice of the objective function in~\secref{s:costFunc}. We solve the optimization problem using a stochastic optimization algorithm called the covariance matrix adaptation evolution strategy (CMA-ES)~\cite{hansen-ost01,muller-kou-e02,hansen-kou-e03,hansen-kern04}. Evaluating the objective function requires simulating cell flows through a DLD device. That's why, it is infeasible to solve the optimization problem using our {\em high-fidelity DLD model (HF-DLD)}. So, we propose a {\em low-fidelity DLD model (LF-DLD)} that has less number of unknowns than HF-DLD. We build HF-DLD once to obtain accurate boundary conditions for LF-DLD and perform the simulations in the LF-DLD. Once the optimization is solved with LF-DLD, the result is verified in HF-DLD. We also carefully analyze the sensitivity of cell dynamics to the perturbations in pillar cross sections using HF-DLD. We consider four sorting scenarios involving cells with similar viscosity contrast values. We compare the features of the optimal designs for these scenarios with the conventional ones which have equal gap sizes and circular, triangular, square and diamond cross sections.

\subsection{Contributions}

To the best of our knowledge, this is the first study investigating optimal DLD designs for deformability-based sorting of RBCs. The main contribution of our study is to show that designing a DLD device can be posed as an optimization problem and solving it systematically discovers optimal designs as opposed to doing an exhaustive search. The optimal designs in the scenarios considered here are different than those in the literature and have cross sections similar to a triangle with a flat edge in the shift direction of the pillars (see~\figref{f:motivation}). These designs are optimal in the sense that they provide large vertical separation between the cells after a few number of cell-pillar interactions. Therefore, they provide efficient cell sorting. The other contribution is a low-fidelity DLD model which reduces the computational cost of the simulations so that an optimization problem can be solved in a reasonable time (i.e., 3-4 days whereas it would take 15-16 days if the high-fidelity model was used.).

\subsection{Limitations}

Our simulations are in two dimensions. We have opted to use two-dimensional simulations since three-dimensional simulations for cell sorting via DLD can be quite expensive for optimization~\cite{rossinelli-petros-e15}. We consider only dilute suspensions. We do not allow changes in the resting size and shape of an RBC. We do not put any constraints on the pillar cross sections regarding the manufacturability such as symmetry of the cross sections. As a result of that, the optimal designs might have sharp and fine features. However, these designs can still help design an effective device with simpler geometry in~\secref{s:design}.

\subsection{Related work}

We refer the reader to~\cite{kabacaoglu-biros17} for the review of related work on deformability-based cell sorting and the details of our numerical scheme for cell flow simulations. Here, we review the literature on DLD designs. 

There are only a few studies considering pillar cross sections different than circular. \citet{loutherback-sturm-e10} proposed a triangular pillar cross section and studied the effects of its size and orientation, vertex rounding on size-based sorting of rigid spherical particles. They found that the triangular pillar cross section shifts the flow in a gap towards the sharp vertex. This reduces the width of the stream adjacent to the sharp vertex compared to a circular cross section. So, for the same adjacent stream size a triangular pillar cross section allows using larger vertical gap size than a circular one, which not only increases the throughput but also reduces the risk of clogging. While a triangular cross section can be used to adjust the critical particle size for the separation of rigid particles, it cannot be straightforwardly used for deformability-based sorting of cells because it also affects cell dynamics which is not investigated in~\citet{loutherback-sturm-e10}. Our optimal designs have cross sections similar to a triangle and we investigate the effects of such cross sections on cell dynamics. In~\cite{loutherback-sturm-e12} the same group used such a design in an experimental study to sort circulating tumor cells from whole blood and proved the advantages of their design. \citet{alfandi-alebbini-e11} aimed at proposing cross sections such that the cells deform only slightly and their dynamics can be predicted using the analytical DLD theory for rigid particles. They did not consider deformability-based sorting of the cells. The proposed cross section has an airfoil shape and results in a velocity field which does not deform cells as much as circular or diamond cross sections. \citet{zeming-zhang-e13} aimed at designing a DLD device in which healthy RBCs displace. They suggested an I-shape cross section, which can induce rotational motion (tumbling) of the cells and hence lead them to stay out of the adjacent stream. They proved the effectiveness of the proposed design by conducting experiments. \citet{ranjan-zhang-e14} experimentally studied the effects of various orientations of I-shape, T-shape and L-shape cross sections on the dynamics of rigid spherical particles and the cells. They concluded that cross sections with protrusions and grooves can induce tumbling of the cells and therefore, lead them to displace while keeping the rigid particles zig-zagging. \citet{zhang-fedosov-e15} conducted a numerical study to investigate cell dynamics in DLD devices with circular, square, diamond and triangular pillar cross sections. They stated that the prediction of the cell transit strongly depends on device geometry and structure. Therefore, they expected that new designs other than the circular pillar cross section can be useful for various objectives. All of these studies considered equal horizontal and vertical gap sizes. Recently, \citet{zeming-zhang-e16} showed by conducting experiments that one can change particles' transport modes by using unequal gap sizes without changing the pillar cross section which was circular in particular. Although none of these studies investigated designs for sorting cells by their deformability, they contribute to our understanding of how various cross sections and unequal gap sizes affect cell dynamics. Our formulation results in cross sections that are different than the above studies and automatically adjusts both cross sections and gap sizes.

\subsection{Organization of the paper}

In~\secref{s:vesicles}, we present the mathematical model and the integral equation formulation for cell flows in DLD devices and introduce our low-fidelity DLD model. We propose an objective function and state the design optimization problem and explain how we solve it in~\secref{s:optimProb}. We, then, perform numerical experiments and discuss the results in~\secref{s:results}. Finally, we illustrate how the solution of the optimization problem guides designing a DLD device with a simpler geometry (thus, possibly easier to manufacture) in~\secref{s:design}. 

\section{Modeling\label{s:vesicles}} 
We state the mathematical models of cell flows and DLD device in~\secref{s:governEqns} and in~\secref{s:dldModel}, respectively. We, then, present the nondimensional parameters in~\secref{s:nondim}.

\begin{figure}[!htb]
\begin{minipage}{0.5\textwidth}
\setcounter{subfigure}{0}
\centering
\renewcommand*{\thesubfigure}{(a)} 
      \hspace{0cm}\subfigure[Simulation domain]{\scalebox{0.6}{{\includegraphics{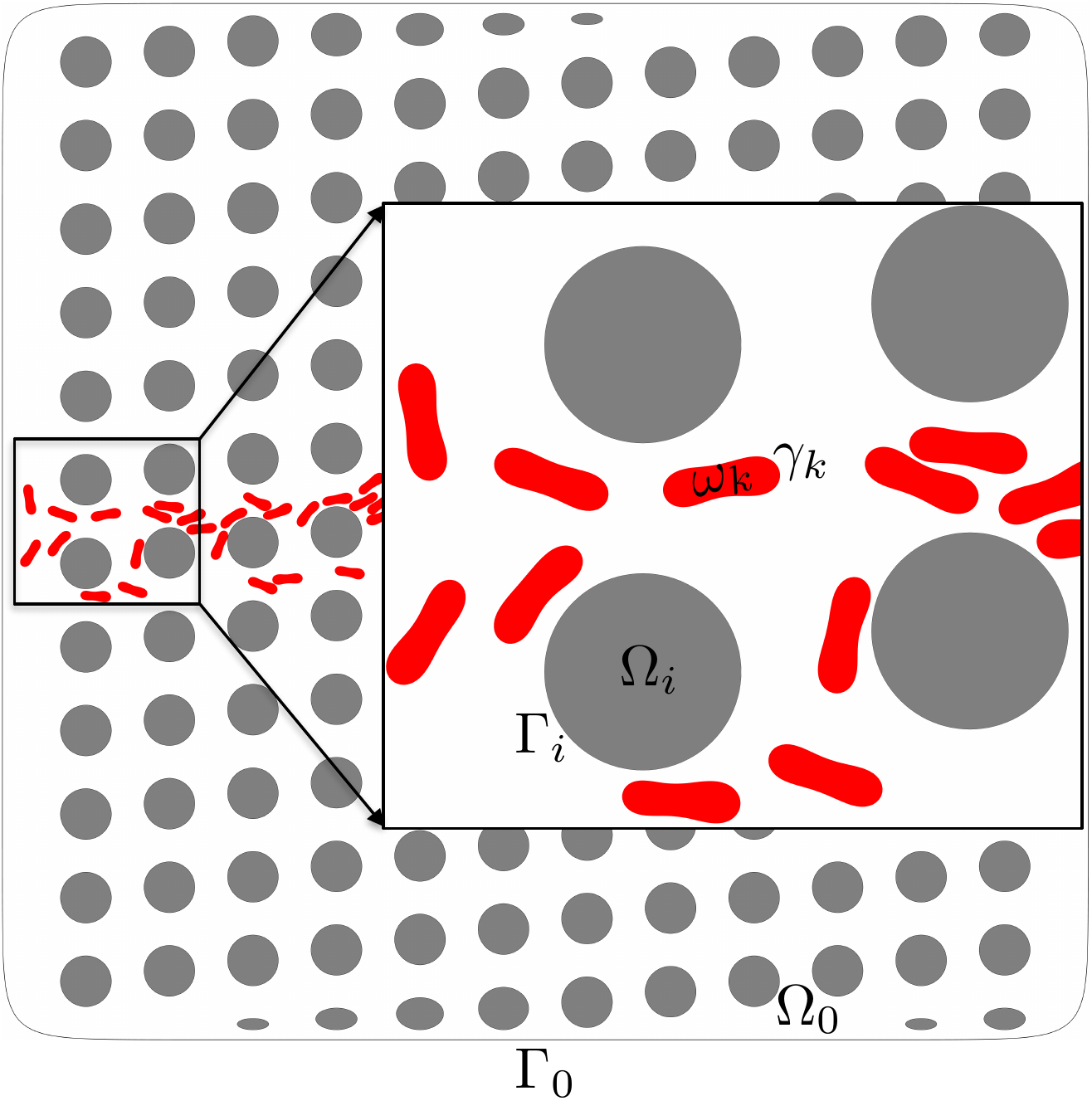}}}
      \label{f:vesInDLD}}
\end{minipage}
\begin{minipage}{0.65\textwidth}
\setcounter{subfigure}{0}      
\centering
\renewcommand*{\thesubfigure}{(b)} 
      \hspace{-1.5cm}\subfigure[Pillar lattice for an arbitrary cross section]{\scalebox{0.65}{{\includegraphics{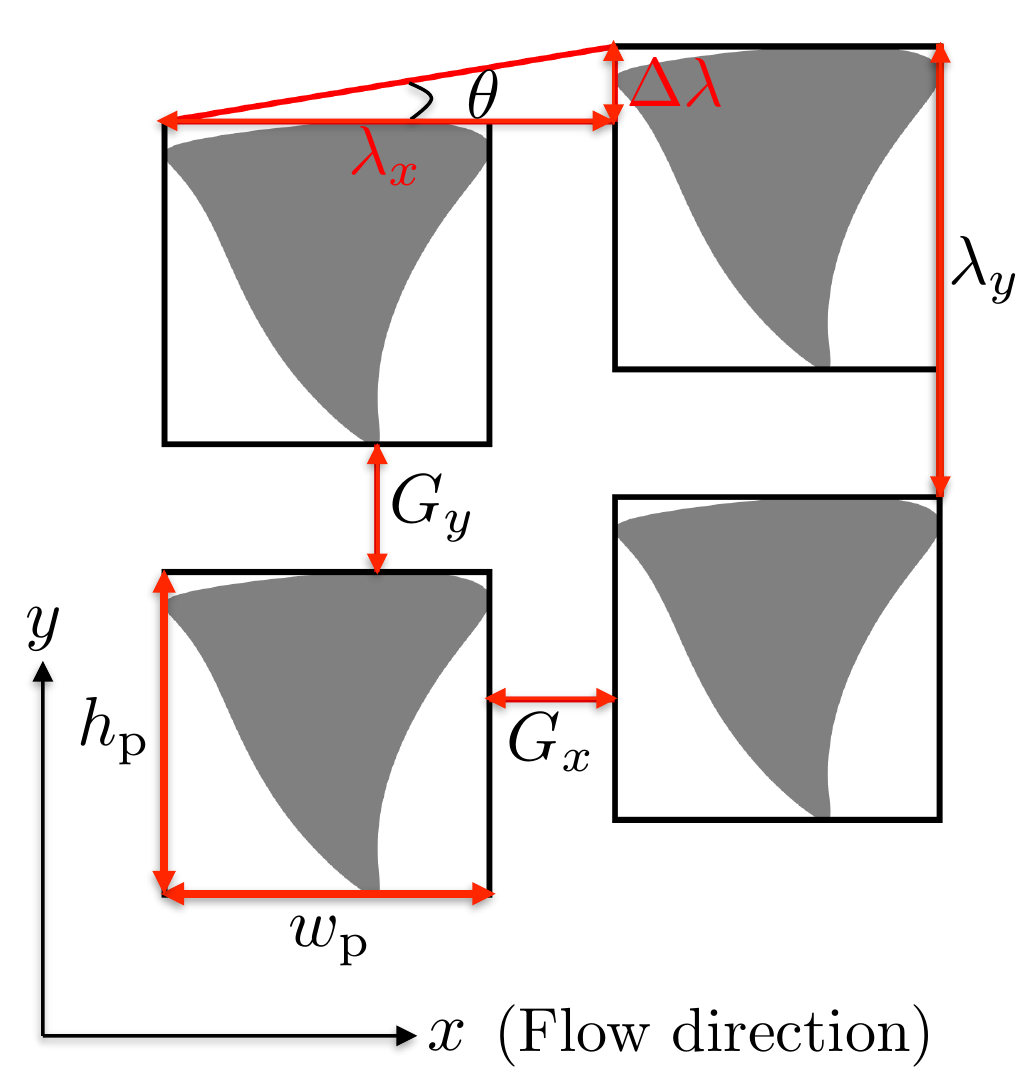}}}
      \label{f:placingPillars}} 
\end{minipage}                    
\mcaption{The domain of a cell flow in a DLD device on the left and top view of the lattice of pillars with arbitrary cross sections (in gray) in a DLD device on the right. On the left, the
interior and boundary of the $i^{th}$ pillar are denoted by $\Omega_i$
and $\Gamma_i$ ($\Omega_0$ and $\Gamma_0$ are the ones of the exterior
wall). Additionally, $\omega_k$ and $\gamma_k$ stand for the interior
and boundary of the $k^{th}$ cell. On the right, $G_x, G_y$ denote the gap sizes, and $\theta$ is the tilt angle of the pillar rows. Fluid flows in the $x$ direction (horizontal) and each pillar column is shifted in the $y$ direction (vertical) by $\Delta \lambda$ with respect to a previous column. We set the gap sizes such that they are the spacings between the circumferential rectangles. $h_\mathrm{p}$ and $w_{\mathrm{p}}$ are the height and the width of the rectangle. The center-to-center distances in the $x$ and the $y$ directions between the rectangles are $\lambda_x = G_x + w_{\mathrm{p}}$ and $\lambda_y = G_y + h_{\mathrm{p}}$, respectively.}{f:explainFigs}
\end{figure}

\subsection{Governing equations}\label{s:governEqns}

We assume that the flow is two-dimensional and there are no external body forces such as gravity. The RBC's interior fluid and the suspending exterior fluid are Newtonian. See~\figref{f:vesInDLD} for the simulation domain. Let the boundary of each pillar be denoted by $\Gamma_i$. The exterior wall $\Gamma_0$ bounds the pillars. $\Omega_0$ and $\Omega_i$ denote the areas enclosed by the exterior wall and the $i^{th}$ pillar, respectively. We define
\begin{equation*}
\Omega = \Omega_0 \setminus \left( \bigcup_i \Omega_i \right),
\end{equation*}
with boundary $\Gamma = \Gamma_0 \cup \left(\bigcup_i \Gamma_i\right)$.

Based on the mechanical properties of RBCs \cite{goldsmith-skalak75,popel-johnson05}, we model them as locally inextensible \editS{vesicles} which can resist bending and tension. The reduced area, $4\pi A/P^2$, is the ratio between the area of a cell $A$ and the area of a circle having the same perimeter $P$. The RBCs in this study have the reduced area $0.65$. They exhibit biconcave shape with the diameter $8\mu m$ and the thickness $3\mu m$. We consider only the same-size cells.

Let $\gamma_k$ and $\omega_k$ stand for the boundary and interior of
the $k^{th}$ cell, respectively. 
Then $\gamma = {\bigcup}_k \gamma_k$ and $\omega = {\bigcup}_k \omega_k$. We consider the DLD applications in low Reynolds number regime (i.e., $\bigO(10^{-3})$)~\cite{mcgrath-bridle-e14} so we assume Stokesian fluids. In that regime momentum and mass conservation are given by
\begin{equation} \label{e:stokesCont}
  -\eta \Lap \uu (\xx) + \Grad p(\xx) = 0, 
  \quad \text{and} \quad \Div(\uu(\xx)) = 0, \quad 
  \xx \in \Omega \setminus \gamma.
\end{equation}
Here, $\eta$ is the viscosity, $\uu$ is the
velocity and $p$ is the pressure. We impose Dirichlet boundary conditions for the velocity on all fixed walls:
\begin{equation}\label{e:noslipOnWalls}
\uu(\xx,t) = \UU (\xx,t), \quad  \xx \in \Gamma.
\end{equation}
Velocity is required to be continuous on the cell interface, i.e.,
\begin{equation}\label{e:velContinuity}
\uu(\xx,t) = \frac{d \xx}{dt}(t), \quad  \xx \in \gamma.
\end{equation}
RBCs are known to be nearly inextensible, i.e., they locally conserve  arc-length in 2D. Therefore,
\begin{equation}\label{e:inexten}
\xx_s \mathbf{\cdot} \uu_s = 0, \quad  \xx \in \gamma,
\end{equation}
where the subscript "$s$'' stands for differentiation with respect to
the arc-length on the boundaries of cells. Finally, we impose the momentum balance on the cell interface. Since the cells
resist bending and tension, the interface applies an elastic force as
a result of deformation due to them. The momentum balance enforces the
jump in the surface traction to be equal to the net elastic force
applied by the interface,
\begin{equation} \label{e:tracJump}
  [\![ \mathbf{T}\nn(\xx) ]\!] = -\kappa_b \xx_{ssss} + \left(\sigma(\xx,t) \xx_s \right)_s,  \quad
  \xx \in \gamma,
\end{equation}
where $\mathbf{T} = -p\mathbf{I} + \eta (\Grad \uu + \Grad {\uu}^T)$ is the
Cauchy stress tensor, $\nn$ is the outward normal vector on $\gamma$,
$[\![ \cdot ]\!]$ is the jump across the interface. The right-hand
side is the net force applied by the interface onto the fluid. The
first term on the right-hand side is the force due to bending
stiffness $\kappa_b$ and the second term is the force due to tension
$\sigma$, which acts as a Lagrange multiplier enforcing the
inextensibility \cite{shravan-biros-e09}.  Finally, the position of
the boundaries of $M$ cells evolves as
\begin{equation} \label{e:boundaryEvolve}
  \frac{d\xx_i}{dt} = \uu_{\infty}(\xx_i) + \sum_{j=1}^M \uu(\xx_j), \quad i = 1,\ldots,M, \quad \forall \xx \in \gamma_i,
\end{equation}
where $\uu_{\infty}(\xx_i)$ is the background velocity (imposed by the solid boundaries in this case) and
$\uu(\xx_j)$ is the velocity due to the $j^{th}$ cell acting on the
$i^{th}$ cell. The complete set of nonlinear equations
\eqref{e:stokesCont}-\eqref{e:boundaryEvolve} governs the evolution of
the cell interfaces. In line with our previous work
~\cite{kabacaoglu-biros17}, we use an integral equation method to obtain the positions of the cell boundaries (see~\cite{kabacaoglu-biros-e18} for the details of the numerical scheme). Throughout all calculations we use the same spatial and temporal resolutions. 

\paragraph*{Remark}Let us also define a cell's inclination angle as the angle between the flow direction and the principal axis corresponding to the smallest principal moment of inertia~\cite{rahimian-biros-e10}. The moment of inertia tensor is
\begin{equation*}\label{e:IAeq}
J = \int_{\omega} \left( |\rr|^2 - \rr \otimes \rr \right)d\xx = \frac{1}{4}\int_{\gamma} \rr \cdot \nn \left(|\rr|^2I - \rr \otimes \rr \right)ds,
\end{equation*}
where $\rr = \xx - \mathbf{c}$ and $\mathbf{c}$ is the center of the cell. So, for the cell on the \editS{right} in~\figref{f:dldModels}, the inclination angle is $\alpha$. The cell migration depends on its inclination angle and the curvature of the imposed flow. Soft cells migrate more in the vertical direction than stiff cells do since they have higher inclination angles. The curvature of the imposed flow depends on the DLD design. The cell migration is more pronounced in the flows shifted in the vertical direction. Since we cannot predict the cell migration in DLD flows analytically or using simulations in simpler flows~\cite{kabacaoglu-biros17}, it requires simulating the cells in a given device to determine whether they can be sorted. 

\subsection{DLD model}\label{s:dldModel}

\begin{figure}[!htb]
\begin{center}
\includegraphics[scale=0.75]{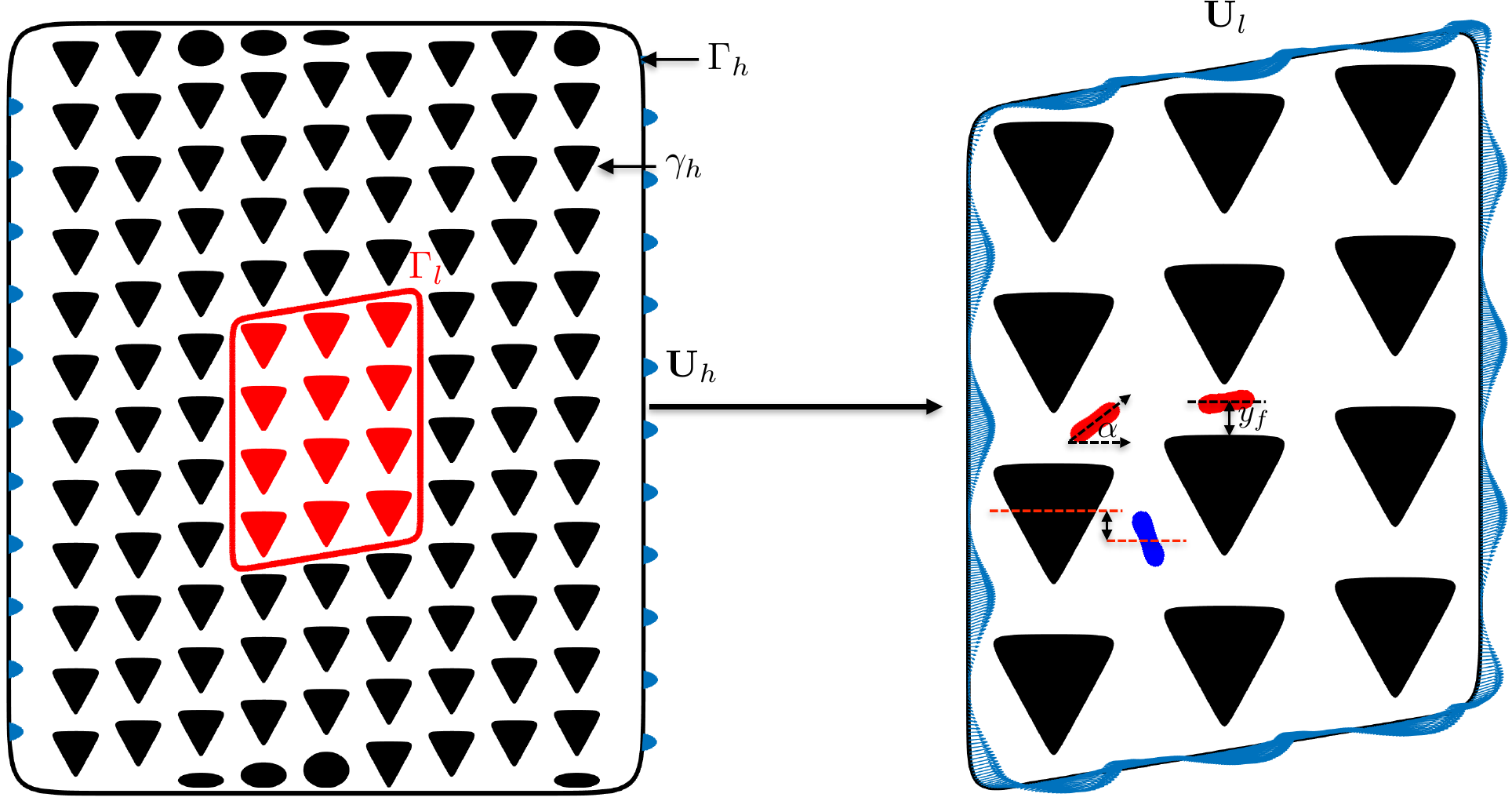} \mcaption{High-fidelity model (on the left) results in large number of unknowns, which renders its use for optimization impractical. We develop a low-fidelity model (on the right) which is constructed as follows. Side walls in the high-fidelity model pass through where the vertical gap size between the pillars is minimum (i.e. $G_y$) on the imaginary columns on the left and on the right. We assign a parabolic velocity $\UU_h$ on the imaginary gaps on $\Gamma_h$ (blue arrows in the left figure) and zero velocity on $\gamma_h$. Then, we solve~\eqref{e:fred2nd} to obtain the density $\zzeta_h$ on $\Gamma = \Gamma_h \cup \gamma_h$. Finally, we compute the velocity $\UU_l$ at $\xx \in \Gamma_l$ due to the density $\zzeta_h$ using~\eqref{e:stokeslet}. We impose $\UU_l$ as a Dirichlet boundary condition when simulating cell flows using the low-fidelity model as in the right figure. \editS{$y_f$ on the right figure is the vertical distance between the displacing cell’s center and the top of the pillar underneath it at the end of the simulation. $\alpha$ is the inclination angle of the cell.}}{f:dldModels}
\end{center}
\end{figure}
\editS{Let us first mentioned the pillar arrangement. The pillars are placed in a DLD device based on the smallest circumferential rectangle (See~\figref{f:placingPillars} for the schematic). Let $h_{\mathrm{p}}$ and $w_{\mathrm{p}}$ be the height and the width of this rectangle. Fluid flows in the $x$ direction, which is the horizontal direction. We denote the horizontal and vertical gap sizes between the rectangles with $G_x$ and $G_y$, respectively. The gap sizes are also the minimum spacings between the pillars. The horizontal and the vertical center-to-center distances between the rectangles become $\lambda_x = G_x + w_{\mathrm{p}}$ and $\lambda_y = G_y + h_{\mathrm{p}}$. Each column is shifted in the vertical direction with respect to the previous column by $\Delta \lambda$ which is defined as $\Delta \lambda = \tan(\theta) \lambda_x$ for the tilt angle of the pillar rows $\theta$. The tilted pillar row arrangement divides the flow in a vertical gap into a several streams carrying equal flux. After every vertical gap the stream adjacent to a pillar ({\em the adjacent stream}) swaps a lane by moving downwards. The number of these streams is $\left \lceil n \right \rceil$, where 
\begin{equation}\label{e:period}
n = \frac{\lambda_y}{\Delta \lambda}. 
\end{equation}
$n$ is also referred to as the number of columns in a period of the device if $n$ is an integer. "Period'' sets a length scale in which the column arrangement is exactly repeated. If $n$ is not an integer, the column arrangement does not repeat itself.} 

In addition to the modeling assumptions on the fluid flow, we also need to introduce an approximation for the device, in particular, boundary conditions. Actual DLD devices usually consist of $\bigO(10)$ tilted rows and $\bigO(100)$ columns, which results in $\bigO(1000)$ pillars in a device. Performing simulations of cell sorting in a whole device is computationally very expensive even for a single simulation, let alone for optimization. In order to evade the computational cost the numerical studies for the simulation of DLD reduced the simulation domain to a single pillar by assuming periodic boundary conditions~\cite{quek-chiam-e11,ye-yu-e14,kruger-coveney-e14,vernekar-kruger15,zhang-fedosov-e15}. Such boundary conditions are tantamount to imposing artificial vertical pressure difference to enforce no net vertical flow~\cite{davino13,zhang-fedosov-e15}. We avoid introducing such a force by using an exterior wall in our model. However, to reduce the cost we make the device smaller than what it is in practice and this in turn can introduce errors. We discuss two approximate models: a high-fidelity and a low-fidelity. Key parameters in these models are the number of rows (width) and columns (length) and the boundary conditions applied on the exterior wall. To describe the length we use the notion of the \emph{period}, which we introduced in~\eqref{e:period}, and $\left \lceil n \right \rceil$ is the number of columns per period. Let us explain these models.

\paragraph*{\bf High-fidelity model (HF-DLD)} The high fidelity model is based on the model we presented in~\cite{kabacaoglu-biros17}, where we numerically determined that wall effects are negligible if we use 12 rows and $\left \lceil 1.5 n \right \rceil$ columns. However, in the present study we observe that these numbers of rows and columns depend on the gap sizes and the pillar cross sections. That's why, in this study HF-DLD has 12 rows and 9 columns of pillars as in the left figure in~\figref{f:dldModels}. Since wall effects are minimum in the middle of the device, we initialize a cell there. As the cell travels and reaches to the next column, we translate it back to the previous column. This trick is possible since unlike~\cite{kabacaoglu-biros17} we are interested in a single cell in the present study.  So simulations take place between only two columns. Convergence studies for the cell trajectories showed that the wall effects are negligible for various pillar cross sections in HF-DLD. Using this model we can simulate a single cell for an arbitrary number of periods with a much smaller cost. \editS{Initial position and orientation of a cell have to respect the physics of cell flows in DLD. Displacing cells have asymptotic periodic motion with a certain distance to pillars in a gap and positive inclination angles~\cite{kabacaoglu-biros17}. This certain distance depends on the cell's deformability and the flow field. Zig-zagging cells get much closer to pillars and have negative inclination angles. In order to minimize the uncertainty in the initial position and orientation of a cell, we initialize it in the middle of a vertical gap with zero inclination angle.} Although the flow is not symmetric in the gap for arbitrary pillar cross sections, we still impose a symmetric parabolic velocity as a boundary condition ($\UU_h$ in the left figure in~\figref{f:dldModels} as in~\cite{kabacaoglu-biros17}). While this introduces an error, it is negligible in the middle of the device. 

\begin{figure}[!htb]
\begin{minipage}{\textwidth}
\setcounter{subfigure}{0}
\centering
\renewcommand*{\thesubfigure}{(a)} 
      \hspace{0cm}\subfigure[Velocity in HF-DLD]{\scalebox{0.75}{{\includegraphics{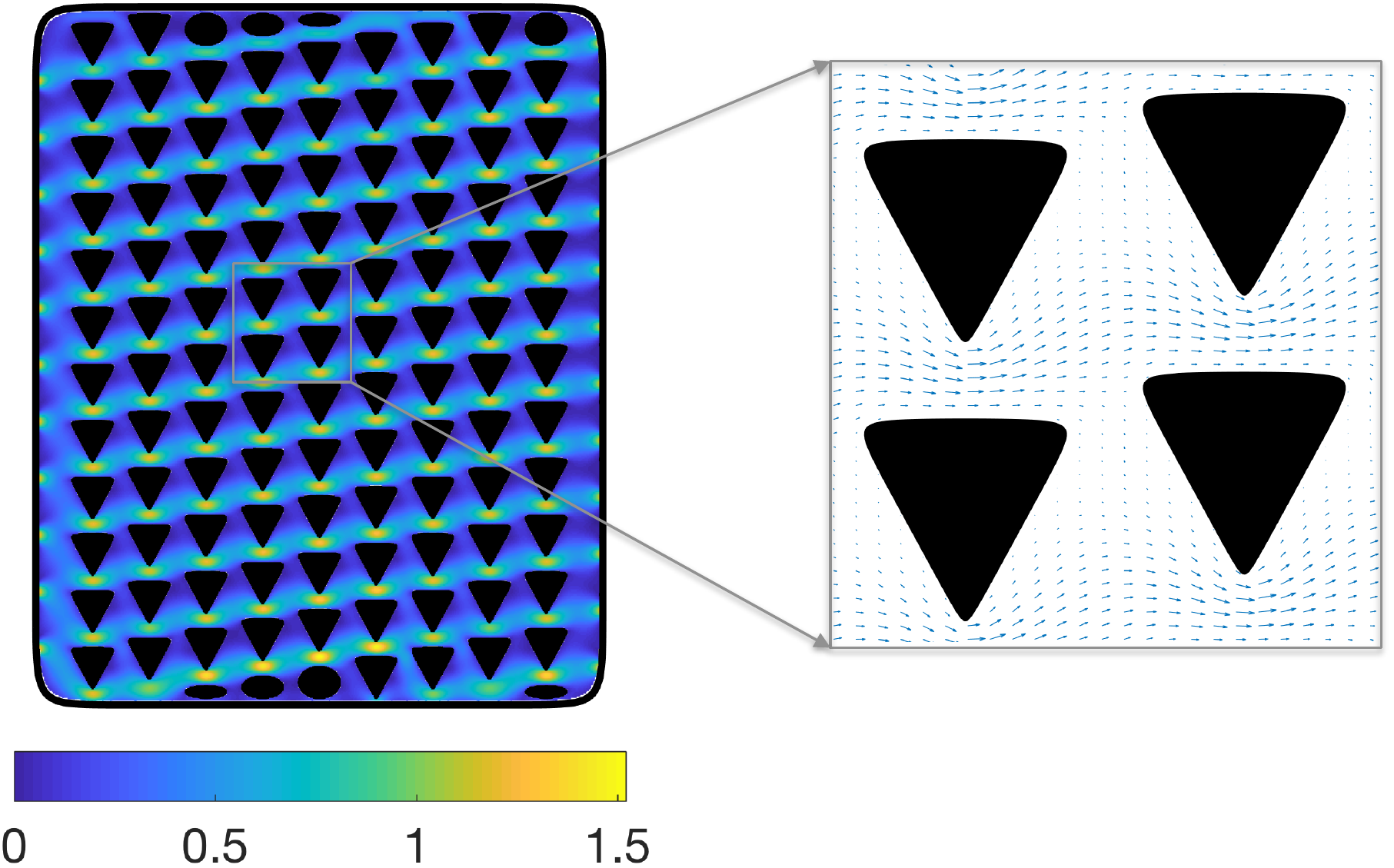}}}
      \label{f:fullDLDvel}} 
\end{minipage}
\begin{minipage}{\textwidth}
\setcounter{subfigure}{0}      
\centering
\renewcommand*{\thesubfigure}{(b)} 
      \hspace{0cm}\subfigure[Velocity in LF-DLD]{\scalebox{0.8}{{\includegraphics{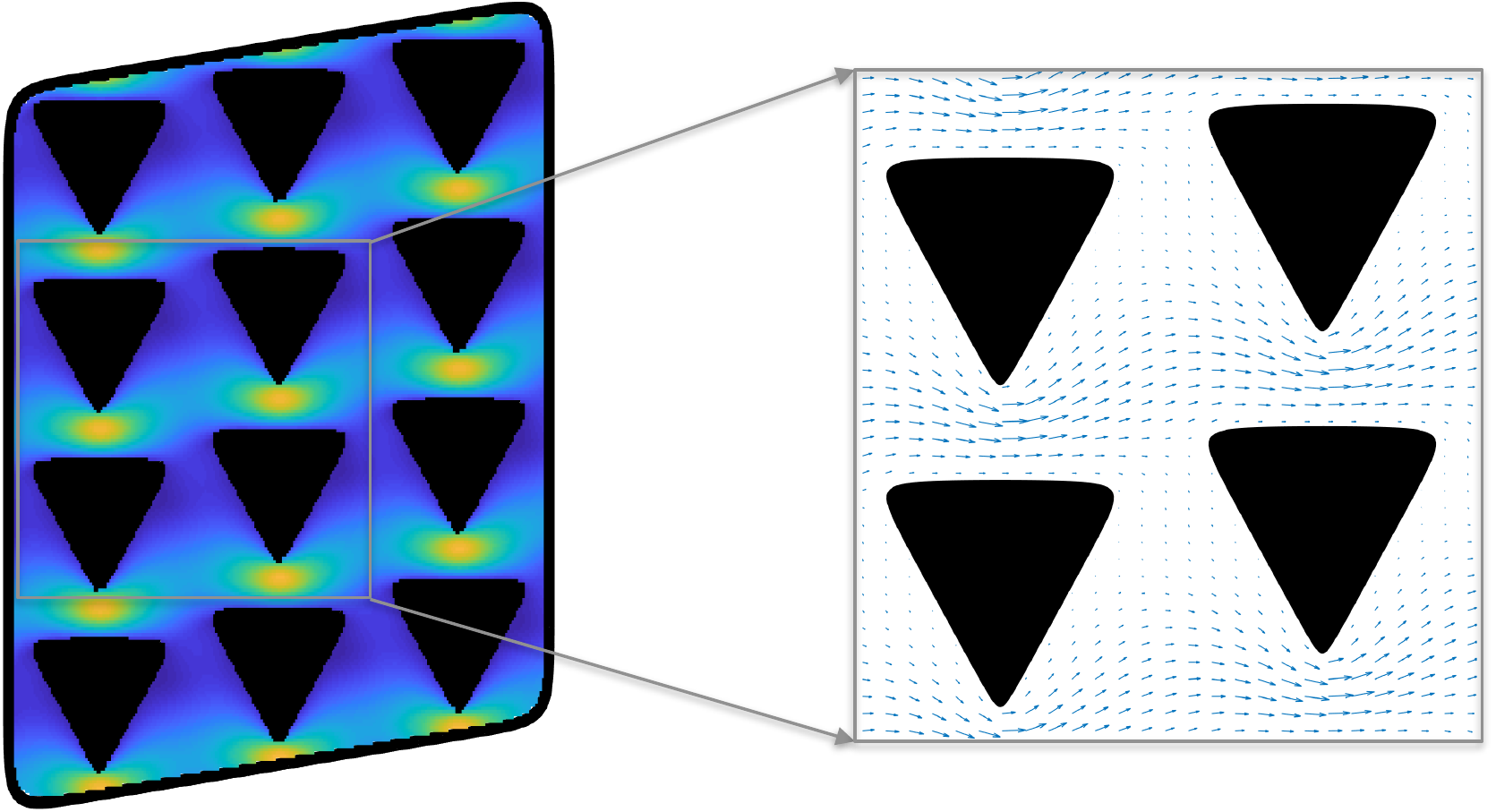}}}
      \label{f:redDLDvel}}
\end{minipage}                    
\mcaption{Velocity magnitude and field in HF-DLD~\figref{f:fullDLDvel} and LF-DLD~\figref{f:redDLDvel} without RBCs for a triangular pillar cross section. The gap sizes are $G_x = G_y = 7.5\mu m$. The dimensions of triangle are $h_{\mathrm{p}} = w_{\mathrm{p}} = 17.5 \mu m$.}
{f:velInDLDmodels}
\end{figure}

\paragraph*{\bf Low-fidelity model (LF-DLD)} Although HF-DLD is still much cheaper than a whole device with hundreds of columns, it is still too expensive for optimization. To further reduce the cost we introduce a low-fidelity model that has four rows and three columns along with an exterior wall (see the right figure in~\figref{f:dldModels}). To make LF-DLD more realistic we place the exterior wall as if it passes in the middle of the gaps between the pillars (i.e. $\Gamma_l$ in the left figure in~\figref{f:dldModels}). Then, we use the velocity field from HF-DLD (without RBCs) as a boundary condition for the exterior wall in LF-DLD. We do this as follows (see also~\figref{f:dldModels} for the schematic):
\begin{itemize}
\item Let $\Gamma_h$ and $\Gamma_l$ denote the exterior walls in HF-DLD and LF-DLD, respectively. Also $\gamma_h$ denotes the boundary of the pillars in HF-DLD. 

\item We impose $\UU_h$ as the velocity on $\Gamma_h$ and zero velocity on $\gamma_h$. We solve the second-kind Fredholm integral equation~\eqref{e:fred2nd} for the hydrodynamic density $\zzeta_h$ on the boundary $\Gamma = \Gamma_h \cup \gamma_h$~\cite{pozrikidis92}:
\begin{equation}\label{e:fred2nd}
\UU_h(\xx) = -\frac{1}{2}\zzeta_h(\xx)+\frac{1}{\pi}\int_{\Gamma} \frac{\rr \cdot \nn}{\|\rr\|^2}\frac{\rr \otimes \rr}{\| \rr \|^2}\zzeta_h(\yy)ds_y, \quad \xx \in \Gamma
\end{equation}
where $\rr = \xx - \yy$. 

\item Using~\eqref{e:stokeslet} we compute the velocity $\UU_l$ (see right figure in~\figref{f:dldModels}) at the discretization points $\xx \in \Gamma_l$ due to the density $\zzeta_h(\yy)$ with $\yy \in \Gamma$~\cite{pozrikidis92}. 
\begin{equation}\label{e:stokeslet}
\UU_l(\xx) = \frac{1}{\pi}\int_{\Gamma} \frac{\rr \cdot \nn}{\|\rr\|^2}\frac{\rr \otimes \rr}{\| \rr \|^2}\zzeta_h(\yy)ds_y, \quad \yy \in \Gamma.
\end{equation}

\end{itemize} 

The density $\zzeta_l$ due to $\UU_l$ represents the hydrodynamic sources outside $\Gamma_l$ in HF-DLD. Hence, LF-DLD and HF-DLD give the same velocity at any point in $\Gamma_l$ in the absence of RBCs. See~\figref{f:velInDLDmodels} for the velocity magnitudes and fields for HF-DLD and LF-DLD for a triangular pillar cross section. The presence of the cells in HF-DLD would perturb the velocity $\UU_l$ if it was computed at every time step. For a number pillar cross sections, we measured the space-time average of the perturbation and found that it does not exceed 5\%. Therefore, we consider LF-DLD reliable for optimization. So to be clear, this calculation needs to be repeated whenever the pillar cross section changes but it does not need to be repeated within the calculation of the RBC trajectories.



\subsection{Dimensionless numbers}\label{s:nondim}

A cell's deformability depends on its bending stiffness, the viscosity contrast between the fluid in the interior and the exterior, and the imposed shear rate. There are two dimensionless number quantifying the deformability: (i) the capillary number $C_a$ and (ii) the viscosity contrast $\nu$. The capillary number is the ratio of the applied viscous force on the cell to the resistance due to bending deformation. The viscosity contrast is the ratio of the viscosity of the fluid in the interior and the exterior. They are defined as follows:
\begin{equation*} \label{e:capNumber}
  C_a = \frac{U_{\max}}{G_y/2}\frac{\eta R^3_{\mathrm{eff}}}{\kappa_b}, \quad \text{and} \quad \nu = \frac{\eta_{\mathrm{in}}}{\eta_{\mathrm{out}}}.
\end{equation*}
In the definition of $C_a$, $U_{\max}/({G_y/2})$ is the imposed shear rate in the vertical gap between two pillars, $R_{\mathrm{eff}}$ is the effective radius of the cell ($R_{\mathrm{eff}} = \sqrt{A/\pi}$ where $A$ is the area enclosed by the cell). In the definition of $\nu$, $\eta_{\mathrm{in}}$ and $\eta_{\mathrm{out}}$ are the viscosities of the fluids in the interior and exterior of the cell, respectively. The deformability is proportional to the capillary number and inversely proportional to the viscosity contrast. For a healthy red blood cell in the DLD flows the capillary number is $C_a \in [0.0375, \, 375]$ and healthy young RBCs have viscosity contrast values $\nu \in [4, 6]$~\cite{suresh-seufferlein-e05,popel-johnson05,tomaiuolo14,mcgrath-bridle-e14}. Diseases such as sickle cell anemia, malaria and diabetes result in stiffer RBCs and the bending stiffness increases 10-fold~\cite{suresh-seufferlein-e05}. In all of our simulations we only change the viscosity contrast. We fix the capillary number to $C_a = 3.75$ which corresponds to a value for a healthy RBC flowing through DLD with an average velocity of 1mm/s. This flow speed is in the range [$1\mu$m/s, 10 mm/s] which the DLD experiments considered~\cite{mcgrath-bridle-e14}. If $C_a$ gets larger, then the cells deform significantly and separation by viscosity contrast becomes no longer possible.

\section{Design optimization problem\label{s:optimProb}} 

For given two different viscosity contrast values, we seek a DLD design that sorts the cells by their viscosity contrasts. In~\secref{s:shapeParam} we explain the device parameterization. Then, we state the optimization problem and propose an objective function in~\secref{s:costFunc}. 

\subsection{Device parameterization for optimization}\label{s:shapeParam}

Pillar cross section, center-to-center distances between the circumferential rectangles and tilt angle of the pillar rows are free design parameters. \editS{We seek designs that are small and result in certain vertical displacement between the cells, which provides efficient sorting. This amounts to fixing the tilt angle and the device size. In the optimization problem we fix the center-to-center distances and the tilt angle since the reported experiments~\cite{mcgrath-bridle-e14} and the numerical studies on sorting RBCs using DLD give information about the pillar arrangement, not the device size.} That is, $\lambda_x$, $\lambda_y$ and $\theta$ in~\figref{f:explainFigs} are fixed. Thus, the only free design parameter is pillar cross section which we parameterize using uniform $5^{th}$ order B-splines \editS{(See \appref{a:bSplineCoeffs} for the details)}. For any cross section with size $h_{\mathrm{p}}$ and $w_{\mathrm{p}}$, we deduce the horizontal and the vertical gap sizes from the center-to-center distances between the circumferential rectangles, i.e. $G_x = \lambda_x - w_{\mathrm{p}}$ and $G_y = \lambda_y - h_{\mathrm{p}}$. With that we have all the design parameters to construct a DLD device. So to be clear, a DLD design involves a pillar cross section, center-to-center distances (or gap sizes) and a tilt angle.

In the optimization, we have to make sure that the velocity fields between different optimization iterations are consistent. To this end, for each proposed pillar cross section, we adjust the velocity boundary conditions so that the imposed total flow rate is the same for all designs. We call DLD designs invalid if they have self-intersecting or rough cross sections. Additionally, small gap sizes lead to large velocity and large pressure drop and increase the risk of clogging. The experimental studies reported so far have used gap sizes greater than $7 \mu m$~\cite{mcgrath-bridle-e14}. Here, we set the minimum gap size allowed to $7 \mu m$ and call a design invalid if it has gap sizes smaller than that. We decide whether a cross section is rough using the decay of the \editS{energy} spectrum of the cross section. \editS{If the high-frequency energy exceeds the low-frequency energy, then we consider that cross section rough.} Those invalid configurations are rejected by assigning a high default objective function value.

\editS{\paragraph*{Remark} It is also possible to optimize the tilt angle and the separation between pillars in addition to the pillar cross section. This does not pose any numerical challenge and it would be even easier to find a design that can sort the cells. Here, we aim at optimizing designs for more difficult cases (i.e., under constraints and for very similar cells).}

\subsection{Optimization problem}\label{s:costFunc}

Given viscosity contrast values of two cells ($\nu_1$, $\nu_2$), center-to-center distance between the circumferential rectangles ($\lambda_x$, $\lambda_y$) and tilt angle of the pillar rows $\theta$, we find an optimal design by 
\begin{enumerate}
\item choosing a design,

\item performing simulations of the cells with the viscosity contrast values ($\nu_1$, $\nu_2$) using LF-DLD,

\item then evaluating the objective function to compare dynamics and decide if the design is acceptable.
\end{enumerate}
In order to choose a design systematically we use the {\em covariance matrix adaptation-evolutionary strategy (CMA-ES)}~\cite{hansen-ost01,muller-kou-e02,hansen-kou-e03,hansen-kern04} as an optimization algorithm. It samples designs from a Gaussian distribution which is updated based on the evaluations of the objective function for the sampled designs in the course of the optimization. The only adjustable parameter of the CMA-ES is the number of samples used in an iteration to update the Gaussian distribution. We set this number to 32 and using a 32-core processor we perform embarrassingly parallel cell simulations to evaluate the objective function. This enables fast solution of the optimization problem. We terminate the iterations when the overall standard deviation decreases below 0.05 or becomes stationary for a few iterations.  

We, now, propose an objective function which compares cell dynamics and assesses whether a design provides efficient cell sorting. Let us introduce the following qualitative definitions that characterize the efficiency of the device:
if both cells are sorted (one displaces one zig-zags) then a device is in \emph{"separation mode"}. Otherwise the device is in \emph{"no-separation mode"}. In order to numerically determine whether a cell displaces or zig-zags we run cell simulations until the cell travels one period of a device, i.e. $\left \lceil n \right \rceil$ columns. Recall that the simulations take place between the first two columns in LF-DLD. If the cell's center is below the center of the pillar above which it is initialized, we call it zig-zagging. For instance, the blue cell in the left figure in~\figref{f:dldModels} zig-zags. Otherwise, we call it displacing. We require the objective function to 
\begin{itemize}
\item give always smaller values for designs in separation mode than for those in no-separation mode because only separation mode is desirable,

\item quantify the difference in cell dynamics for both modes, i.e., distinguish between two designs in separation mode and similarly for those in no-separation mode, 

\item in particular, for distinguishing devices in separation mode decrease when the displacing cell migrates more in the vertical direction or the zig-zagging cell zig-zags earlier because this increases the net vertical separation between the cells and hence provides efficient sorting by reducing the process time.
\end{itemize}
Based on these considerations, we define the following objective function
\begin{equation}\label{e:costFunc}
f = \begin{cases} 
-C_1\frac{y_f}{G_y} + C_2\frac{n_{\mathrm{zz}}}{\left \lceil n \right \rceil}, & \text{for separation mode,} \\
C_3 - |\frac{\Delta y_f}{G_y}| & \text{for no-separation mode where both cells displace}, \\
C_3 - |\frac{\Delta n_{\mathrm{zz}}}{\left \lceil n \right \rceil}| & \text{for no-separation mode where both cells zig-zag,}\end{cases}
\end{equation}
where $C_1, C_2, C_3$ are positive constant coefficients, $y_f$ is the vertical distance between the displacing cell's center and the top of the pillar underneath it at the end of the simulation (see the left figure in~\figref{f:dldModels}), $n_{\mathrm{zz}}$ is the number of columns after which the zig-zagging cell zig-zags for the first time. $\Delta (q)$ stands for the difference between the values quantity $q$ of one cell and the other cell. Let us interpret~\eqref{e:costFunc}. 
\begin{itemize}

\item We normalize the vertical displacement of the displacing cell $y_f$ by the vertical gap size $G_y$, $\frac{y_f}{G_y}$, which tells how much the cell migrates away from a pillar. As the cell goes away from the pillar to the middle of the vertical gap, it travels faster and the sorting becomes quicker.

\item For separation mode, $f$ is the difference between the normalized vertical displacement of the displacing cell ($\frac{y_f}{G_y}$) and the normalized number of columns after which the zig-zagging cell zig-zags ($\frac{n_{\mathrm{zz}}}{\left \lceil n \right \rceil}$). It decreases if the displacing cell migrates more in the vertical direction or the zig-zagging cell zig-zags after a less number of columns. Therefore, a design is more efficient if it results in smaller $f$.

\item $C_1$ and $C_2$ assign different importance on the degree of separation and on how fast separation takes place. Our \editS{simulations} showed that the value of $\frac{y_f}{G_y}$ is usually about $0.2$. However, the ratio $\frac{n_{\mathrm{zz}}}{\left \lceil n \right \rceil}$ is close to 1. In order to make these ratios comparable, we use $C_1 = 1$ and $C_2 = 0.2$. 

\item For no-separation mode, $f$ quantifies how different cell dynamics are. We quantify this difference as either the difference in the vertical displacement of the displacing cells or the difference in the number of columns after which the zig-zag occurs. The more this difference is, the more possible it is to separate cells. So, we want to maximize it. That is, we minimize $-|\frac{\Delta y_f}{G_y}|$ for the displacing cells and $-|\frac{\Delta n_{\mathrm{zz}}}{\left \lceil n \right \rceil}|$ for the zig-zagging cells. 

\item The coefficient $C_3$ is chosen to ensure that separation mode results in a smaller objective function than no-separation mode. Running a few \editS{simulations} we found that $C_3 = 10$ is sufficient. 
\end{itemize}
Overall, we state the design optimization problem as 
\begin{equation*} \label{e:optimProb}
\begin{aligned}
\text{minimize} \quad f\left(c \right) \quad & \text{such that} \,\,  G_x(c), G_y(c) \geq G_{\min} = 7\mu m \\
 & \text{and $\mathbf{X}(c)$ is smooth and not self-intersecting,}
 \end{aligned}%
\end{equation*}
where $f$ is in~\eqref{e:costFunc}, $c$ is the coordinates of the B-spline control points and $\mathbf{X}$ is a pillar cross section.

\paragraph*{Remark} We need the objective function to be discontinuous for the following reason. We are interested in designs in separation mode only, however, we want those in no-separation mode to inform the optimizer for faster convergence. That's why, we do not reject designs in no-separation mode by assigning a high default objective function value. Instead, we make the objective function continuously change among the designs in separation mode and among those in no-separation mode but has a jump between these modes. As a result of that, it can distinguish two designs in separation mode and in no-separation mode. \editS{One can also use an overall measure of vertical separation as an objective function, e.g., the difference in the angle at which the cells migrate on average. Such an objective function is insensitive to how much a displacing cell migrates from the pillar and how early a zig-zagging cell zig-zags. However, we seek designs that result in more migration of the displacing cell and earlier zig-zag of the zig-zagging cell, which provides fast sorting. That's why, we decide to use~\eqref{e:costFunc} as an objective function.}

\section{Numerical experiments\label{s:results}} 
\subsection{Setup}\label{s:setup}
We considered four sorting scenarios with different viscosity contrast values $\nu_1$, $\nu_2$ and tilt angles $\theta$. The details are as follows.
\begin{itemize}

\item \textbf{Scenario 1}: ($\nu_1 = 5$, $\nu_2 = 10$) and $\theta = 0.17$ rad. A healthy RBC has viscosity contrast $\nu \in [4, 6]$~\cite{mohandas-gallagher08}. So, we consider a healthy cell and a cell slightly stiffer than that. In this and the fourth scenarios, the viscosity contrasts are the closest compared to the other scenarios. So, cell dynamics are the most similar and hence it is difficult to separate these cells. The critical viscosity contrast value for separation must be between 5 and 10. The purpose of this experiment is to see if it is possible to design a device to sort cells with very similar dynamics.

\item \textbf{Scenario 2}: ($\nu_1 = 4$, $\nu_2 = 10$) and $\theta = 0.17$ rad. We keep the viscosity contrast of the stiff cell the same as the previous scenario and make the soft cell a little softer. If there is an optimal device for the previous scenario, it sorts the cells in this scenario as well since that device must have the critical viscosity contrast value for separation between 5 and 10. Here, we aim to investigate how much the optimal design for the previous scenario changes due to a slight change in the cells' viscosity contrasts. 

\item \textbf{Scenario 3}: ($\nu_1 = 5$, $\nu_2 = 50$) and $\theta = 0.17$ rad. This is the easiest scenario since the viscosity contrast of the stiff cell is 10 times greater than the soft cell. So optimal devices for the previous scenarios can sort the cells in this scenario as well. Here, the stiff cell cannot migrate as much as the stiff cells in the previous scenarios due to its greater viscosity contrast value. Our goal is to observe how this fact affects the optimal design.

\item \textbf{Scenario 4}: ($\nu_1 = 5$, $\nu_2 = 10$) and $\theta = 0.2$ rad. We consider the same cells as in scenario 1 but we set the tilt angle to a greater value than scenario 1. The size of the adjacent stream increases with the tilt angle and hence the soft cell has to migrate more to displace. Separating the cells is more difficult in this scenario than scenario 1. So we want to inquire if it is still possible to find a design to separate these cells. 

\end{itemize} 

In all scenarios we fix $\lambda_x = \lambda_y = 25\mu m$ because these are the typical dimensions for a DLD device for rigid or deformable particles~\cite{mcgrath-bridle-e14}. We impose the horizontal flux of $5225 (\mu m)^2/s$ in a vertical gap, which sets the capillary number to $C_a = 3.75$ as discussed in~\secref{s:nondim}. We set the tilt angle of the pillar rows to $\theta = 0.17\,\,\mathrm{rad}$ in the first three scenarios and to $\theta = 0.2\,\,\mathrm{rad}$ in the last scenario.  These tilt angles are relatively large compared to the typical angles used in the real devices~\cite{mcgrath-bridle-e14}. We have chosen large angles because they are more challenging (separation is harder to achieve) but also more desirable (if separation is possible, it is faster since shorter devices can be used). 

We use the same initial guess for all scenarios. The initial guess has a circular cross section with the diameter $15\mu m$ and the gap sizes are $G_x = G_y = 10 \mu m$\footnote{Many other experimental and computational studies mentioned in~\secref{s:intro} have considered similar design parameters.}. For all the scenarios that we described above this design does not sort the cells (i.e. in no-separation mode)~\cite{kabacaoglu-biros17}. All the numerical experiments are done in MATLAB on 32-core 2.60GHz Intel Xeon processor with 256GB memory.

\subsection{Results and discussions}

After a number of iterations the optimization algorithm was able to find optimal designs in separation mode for all the scenarios. \editS{These designs are optimal only for the scenarios, the objective function we defined in~\eqref{e:costFunc} and under the size constraints.} We, first, present and discuss the optimal DLD designs in~\secref{s:optimalDesigns}. Then, we show the results of the high-fidelity model simulations using the initial guess and the optimal designs in~\secref{s:fullModelSims}. 

\subsubsection{Optimal designs}\label{s:optimalDesigns}

\begin{figure}[!htb]
  \begin{minipage}{\textwidth}
    \begin{center}
    \includegraphics[scale=0.72]{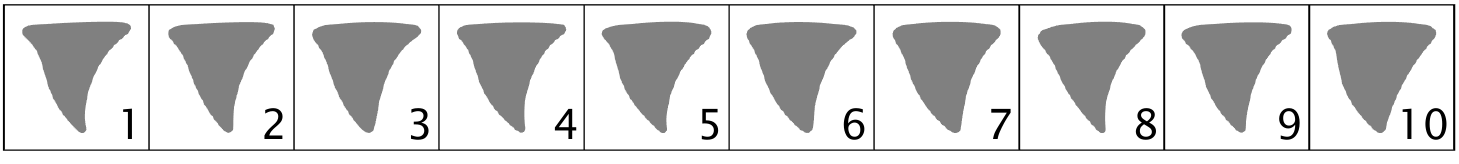} \mcaption{Scenario 1: $(\nu_1, \nu_2) = (5, 10)$ and $\theta = 0.17 \,\, \mathrm{rad}$ (See also~\tabref{t:rest10VCs5a10}).}{f:best10VCs5a10}
    \end{center}
\end{minipage}
  \hfill
\begin{minipage}{\textwidth}
  \centering
    \begin{tabular}{| c | c | c | c | c | c | c | c | c | c | c | }\hline
         \textbf{Design}       & \textbf{1} & \textbf{2} & \textbf{3} & \textbf{4} & \textbf{5} & \textbf{6} & \textbf{7} & \textbf{8} & \textbf{9} & \textbf{10}  \\ \hline
        $G_x (\mu m)$ & 7.5 & 7.8 & 7.4 & 7.3 & 7.3 & 7.3 & 7.6 & 7.7 & 7.3 & 7.5 \\ \hline
        $G_y (\mu m)$ & 7.0 & 7.1 & 7.1 & 7.2 & 7.0 & 7.1 & 7.1 & 7.0 & 7.2 & 7.1 \\ \hline 
        $y_f/G_y$ & 0.356 & 0.380 & 0.378 & 0.376 & 0.349 & 0.372 & 0.371 & 0.370 & 0.369 & 0.344 \\ \hline
        $n_{\mathrm{zz}}$ & 3 & 4 & 4 & 4 & 3 & 4 & 4 & 4 & 4 & 3 \\ \hline
        
      \end{tabular}
      \captionof{table}{\textit{Gap sizes and cell dynamics for designs with cross sections in~\figref{f:best10VCs5a10}.}}
      \label{t:rest10VCs5a10}
\end{minipage} 
\hfill
\begin{minipage}{\textwidth}
    \begin{center}
    \includegraphics[scale=0.72]{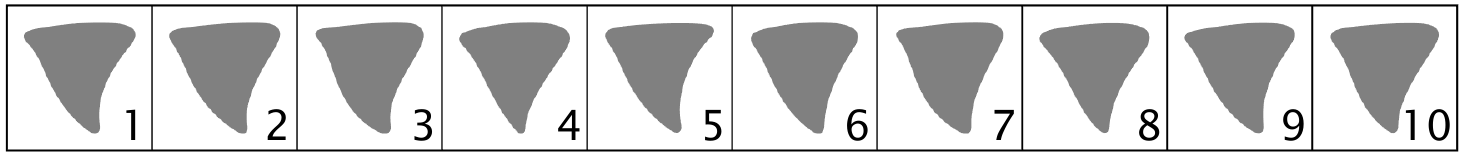} \mcaption{Scenario 2: $(\nu_1, \nu_2) = (4, 10)$ and $\theta = 0.17 \,\, \mathrm{rad}$ (See also~\tabref{t:rest10VCs4a10}).}{f:best10VCs4a10}
    \end{center}
\end{minipage}
  \hfill
\begin{minipage}{\textwidth}
  \centering
    \begin{tabular}{| c | c | c | c | c | c | c | c | c | c | c | }\hline
         \textbf{Design}       & \textbf{1} & \textbf{2} & \textbf{3} & \textbf{4} & \textbf{5} & \textbf{6} & \textbf{7} & \textbf{8} & \textbf{9} & \textbf{10}  \\ \hline
        $G_x (\mu m)$ & 7.0 & 7.1 & 7.5 & 7.1 & 7.4 & 7.7 & 7.8 & 7.3 & 7.3 & 7.6 \\ \hline
        $G_y (\mu m)$ & 7.1 & 7.0 & 7.0 & 7.0 & 7.4 & 7.0 & 7.0 & 7.4 & 7.2 & 7.3 \\ \hline 
        $y_f/G_y$ & 0.377 & 0.402 & 0.375 & 0.374 & 0.398 & 0.370 & 0.370 & 0.394 & 0.369 & 0.394 \\ \hline
        $n_{\mathrm{zz}}$ & 3  & 4 & 3 & 3 & 4 & 3 & 3 & 4 & 3 & 4 \\ \hline
      \end{tabular}
      \captionof{table}{\textit{Gap sizes and cell dynamics for designs with cross sections in~\figref{f:best10VCs4a10}.}}
      \label{t:rest10VCs4a10}
\end{minipage}
\hfill
\begin{minipage}{\textwidth}
    \begin{center}
    \includegraphics[scale=0.72]{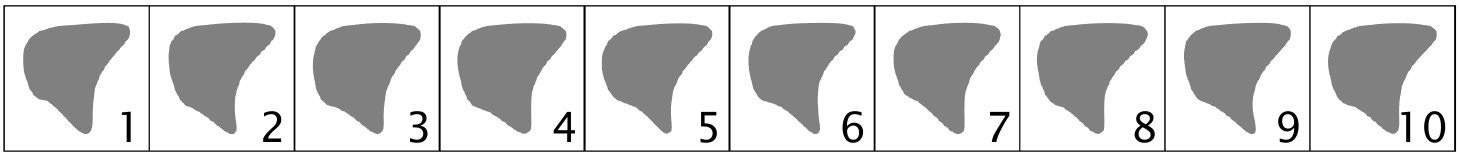} \mcaption{Scenario 3: $(\nu_1, \nu_2) = (5, 50)$ and $\theta = 0.17 \,\, \mathrm{rad}$ (See also~\tabref{t:rest10VCs5a50}).}{f:best10VCs5a50}
    \end{center}
\end{minipage}
  \hfill
\begin{minipage}{\textwidth}
  \centering
    \begin{tabular}{| c | c | c | c | c | c | c | c | c | c | c | }\hline
         \textbf{Design}       & \textbf{1} & \textbf{2} & \textbf{3} & \textbf{4} & \textbf{5} & \textbf{6} & \textbf{7} & \textbf{8} & \textbf{9} & \textbf{10}  \\ \hline
        $G_x (\mu m)$ & 7.7 & 7.8 & 7.6 & 7.5 & 7.3 & 7.8 & 7.7 & 7.1 & 7.7 & 7.5 \\ \hline
        $G_y (\mu m)$ & 7.1 & 7.0 & 7.1 & 7.2 & 7.2 & 7.1 & 7.0 & 7.1 & 7.1 & 7.2 \\ \hline 
        $y_f/G_y$ & 0.367 & 0.364 & 0.363 & 0.362 & 0.362 & 0.361 & 0.361 & 0.360 & 0.360 & 0.359 \\ \hline
        $n_{\mathrm{zz}}$ & 2 & 2 & 2 & 2 & 2 & 2 & 2 & 2 & 2 & 2 \\ \hline
      \end{tabular}
      \captionof{table}{\textit{Gap sizes and cell dynamics for designs with cross sections in~\figref{f:best10VCs5a50}.}}
      \label{t:rest10VCs5a50}
\end{minipage} 
\hfill
\begin{minipage}{\textwidth}
    \begin{center}
    \includegraphics[scale=0.72]{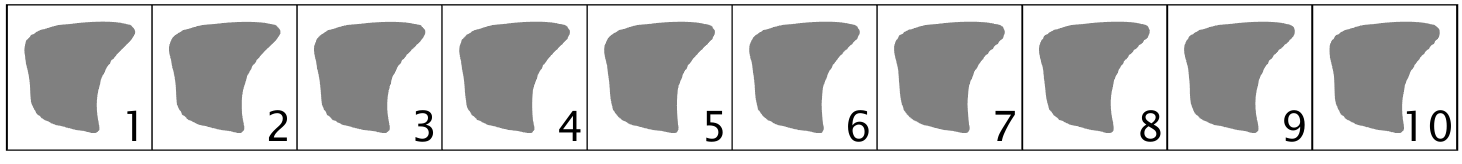} \mcaption{Scenario 4: $(\nu_1, \nu_2) = (5, 10)$ and $\theta = 0.2 \,\, \mathrm{rad}$ (See also~\tabref{t:rest10VCs5a10p5}).}{f:best10VCs5a10p5}
    \end{center}
\end{minipage}
  \hfill
\begin{minipage}{\textwidth}
  \centering
    \begin{tabular}{| c | c | c | c | c | c | c | c | c | c | c | }\hline
         \textbf{Design}       & \textbf{1} & \textbf{2} & \textbf{3} & \textbf{4} & \textbf{5} & \textbf{6} & \textbf{7} & \textbf{8} & \textbf{9} & \textbf{10}  \\ \hline
        $G_x (\mu m)$ & 7.1 & 7.0 & 7.0 & 7.1 & 7.1 & 7.2 & 7.1  & 7.1 & 7.2 & 7.2 \\ \hline
        $G_y (\mu m)$ & 7.0 & 7.1 & 7.1 & 7.0 & 7.0 & 7.0 & 7.0 & 7.0 & 7.0 & 7.0 \\ \hline 
        $y_f/G_y$ & 0.371 & 0.370 & 0.370 & 0.370 & 0.370 & 0.370 & 0.370 & 0.370 & 0.369 & 0.369 \\ \hline
        $n_{\mathrm{zz}}$ & 3 & 3 & 3 & 3 & 3 & 3 & 3 & 3 & 3 & 3 \\ \hline
      \end{tabular}
      \captionof{table}{\textit{Gap sizes and cell dynamics for designs with cross sections in~\figref{f:best10VCs5a10p5}.}}
      \label{t:rest10VCs5a10p5}
\end{minipage}
\end{figure}

\begin{figure}[!htb]
\begin{minipage}{0.5\textwidth}
\setcounter{subfigure}{0}
\centering
\renewcommand*{\thesubfigure}{(a)} 
      \hspace{-0.2cm}\subfigure[Transport modes in the optimal designs]{\scalebox{0.68}{{\includegraphics{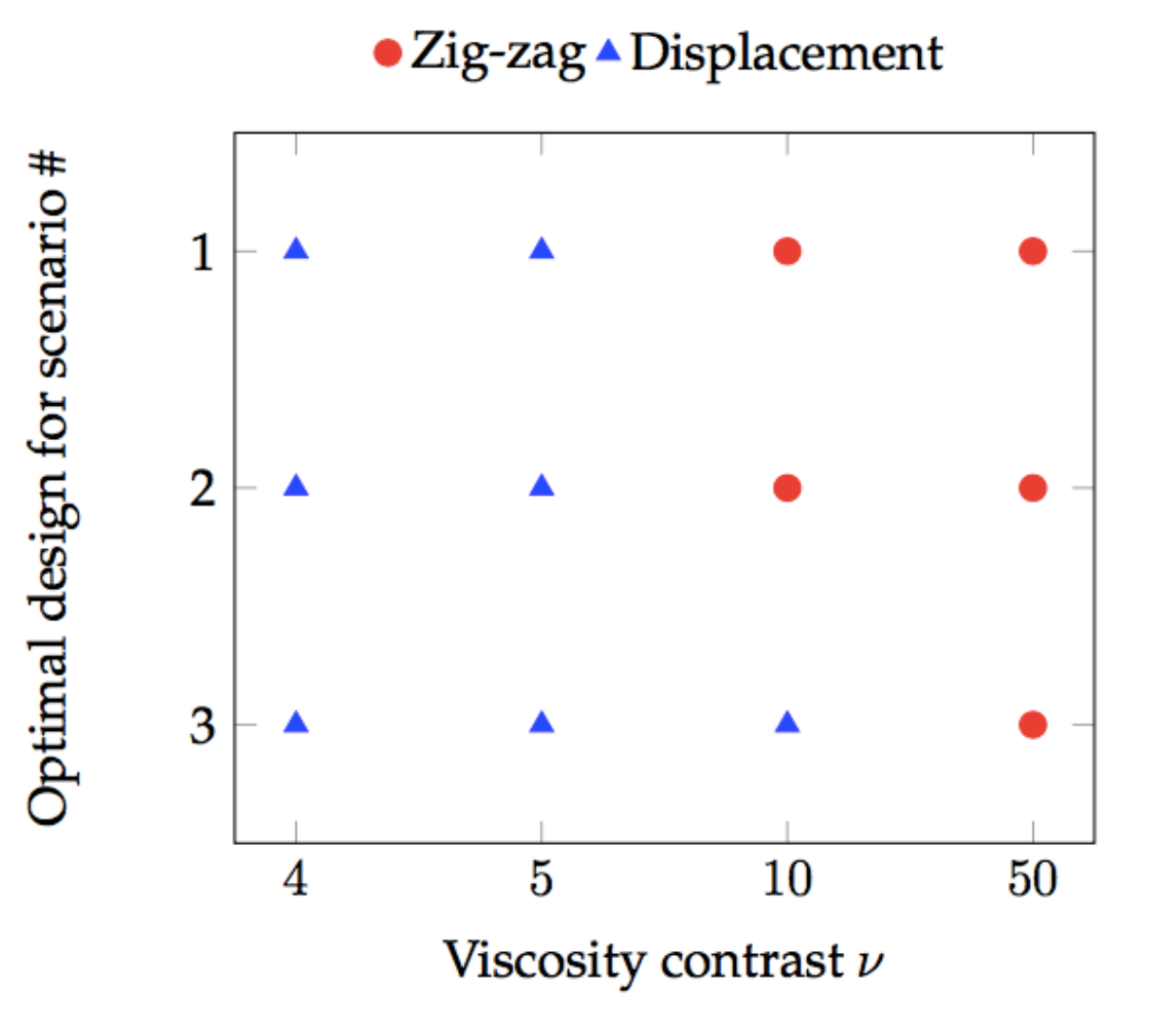}}}
      \label{f:optimalVsVCs}} 
\end{minipage}
\begin{minipage}{0.5\textwidth}
\setcounter{subfigure}{0}      
\centering
\renewcommand*{\thesubfigure}{(b)} 
      \hspace{0cm}\subfigure[Separation mode of designs with circular pillars]{\scalebox{0.68}{{\includegraphics{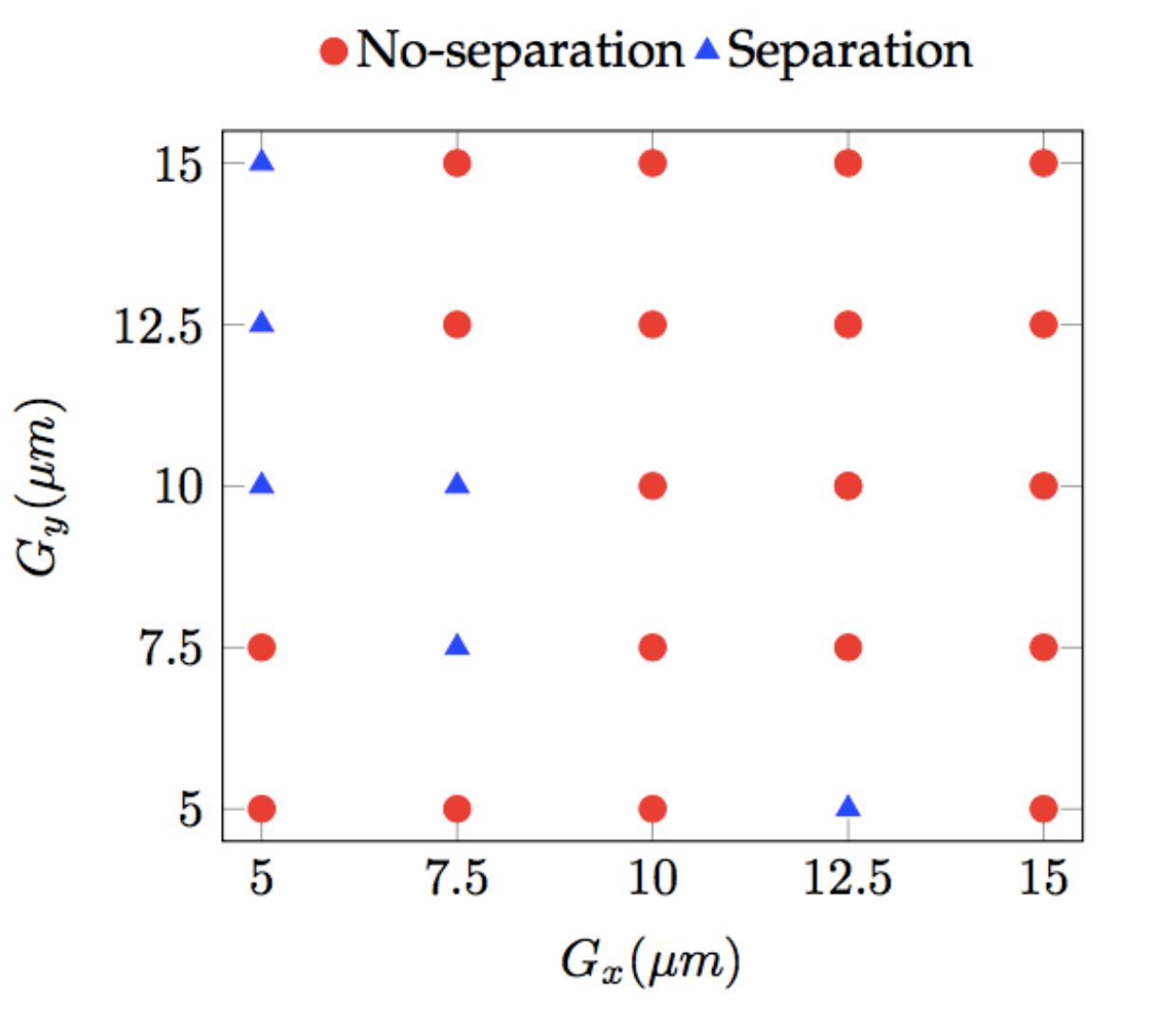}}}
      \label{f:exhaustCirc}}
\end{minipage}                    
\mcaption{On the left: phase diagram for the transport modes of the cells in the optimal designs for the scenarios 1, 2 and 3 as a function of viscosity contrast $\nu$. Displacing and zig-zagging cells are demonstrated with a blue triangle and a red circle, respectively. On the right: phase diagram for sorting cells with the viscosity contrasts $\nu_1 = 5$ and $\nu_2 = 10$ using cylindrical pillars as a function of the gap sizes $G_x$ and $G_y$. We indicate pairs of gap sizes leading to separation with a blue triangle and to no-separation with a red circle.}{f:exhaustSearch}
\end{figure}

For each sorting scenario described in~\secref{s:setup} we present the pillar cross sections for the ten most optimal designs in Figures~\ref{f:best10VCs5a10}-\ref{f:best10VCs5a10p5}. Here, the cross sections are ordered such that the first one results in the smallest objective function value \editS{(i.e., the most efficient sorting)} and the squares around the cross sections have dimensions $\lambda_x = \lambda_y = 25 \mu m$. See~\appref{a:bSplineCoeffs} for the coordinates of the control points for B-spline curves and reproduction of the cross sections. We also tabulate the gap sizes $G_x$, $G_y$ in these designs, the vertical displacements of the displacing cells $y_f/G_y$ and the numbers of columns after which zig-zagging occurs $n_{\mathrm{zz}}$ in Tables~\ref{t:rest10VCs5a10}-\ref{t:rest10VCs5a10p5}. \editS{The designs delivering the smallest objective function value result in either earlier zig-zag of the stiff cells (See Tables~\ref{t:rest10VCs5a10} and~\ref{t:rest10VCs4a10}) or larger vertical migration of the soft cells from the pillars (See Tables~\ref{t:rest10VCs5a50} and~\ref{t:rest10VCs5a10p5}), which leads to larger vertical separation of the cells. So, the small objective function values correlate with the large vertical separation of the cells.} We want to compare the critical viscosity contrast values for separation for the optimal designs in the first three scenarios as well. For that purpose, we perform simulations of the cells with $\nu = (4, 5, 10, 50)$ in the optimal designs and present the cells' transport modes in~\figref{f:optimalVsVCs}. We, now, discuss these results based on the questions raised in~\secref{s:setup}.

The initial guess consists of cylindrical pillars and has gap sizes $G_x = G_y = 10 \mu m$. In all sorting scenarios both soft and stiff cells zig-zag in this initial design. A design in separation mode must lead one of the cells to displace by inducing either more vertical migration and thinner adjacent stream or inducing tumbling motion while keeping the other cell zig-zagging. The optimal designs we found lead the softer cells to displace and the stiffer ones to zig-zag in all scenarios. Those designs have smaller gap sizes and hence thinner adjacent streams than the initial guess. If the pillar cross sections had remained circular, this might be sufficient to separate the cells. In fact, we performed an exhaustive search to find gap sizes that induce separation for the circular pillars (see~\figref{f:exhaustCirc} for the results). The design with circular pillars and gap sizes $G_x = G_y = 7.5\mu m$ can sort the cells in the first three scenarios. However, this design is not among the optimal designs we found because the softer cells cannot migrate as much as they do in the optimal designs. So it has a greater objective function value \editS{(i.e., less efficient sorting)} than the optimal designs. 

Let's try to understand why the optimal designs perform better. They have cross sections with a flat edge at the top and a sharp vertex at the bottom. Such cross sections result in an asymmetric flow in the vertical gap as opposed to a symmetric flow induced by a circular cross section~\cite{loutherback-sturm-e10}. The flow is shifted towards the sharp vertex (e.g., see~\figref{f:redDLDvel}). The flow is from left to right for the optimal cross sections in Figures~\ref{f:best10VCs5a10}-\ref{f:best10VCs5a10p5}. Therefore, the flow is shifted upwards. This has two consequences: First, it results in stronger vertical migration due to larger positive flow curvature; and second, in a thicker adjacent stream compared to a circular cross section with the same gap sizes. Vertical migration is desirable since it can lead cells to displace. Overly thick adjacent stream can be problematic since it can prevent softer cells from displacing. The optimal designs avoid that by having narrower vertical gaps, which decreases the width of the adjacent stream. In order to illustrate that the flow shifted upwards induces more migration, consider the following example. We turn the optimal cross section for scenario 1 upside down, i.e., the sharp vertex is at the top and the flat edge is at the bottom. This results in a thinner adjacent stream since the flow is shifted downwards. We, then, perform a simulation of the cells in scenario 1 in this configuration. This design is also capable of cell sorting: while the stiff cell zig-zags, the soft cell displaces. However, the soft cell migrates less in this design ($y_f/G_y = 0.12$) than in the optimal design ($y_f/G_y = 0.36$) due to the flow shifted downwards. So, the designs we found are optimal since they not only induce cell sorting but also lead the soft cell to displace vertically more than any other possible designs. This discussion also shows that adjusting the width of the adjacent stream does not guarantee separation of cells unlike rigid particles, one has to consider cell migration as well. 

Comparing the optimal designs for the first two scenarios illustrates that a small change in the viscosity contrast value leads to visible but not significant changes in the optimal designs. The stiff cells are the same in these scenarios and have $\nu_2 = 10$. The softer cells have $\nu_1 = 5$ and $\nu_1 = 4$ in scenario 1 and 2, respectively. The optimal designs for scenario 2 (see~\figref{f:best10VCs4a10}) have cross sections similar to those for scenario 1 (see~\figref{f:best10VCs5a10}). 

In scenario 3, the soft cell is the same as in scenario 1 but the stiff cell is much stiffer so that it migrates less. The optimal designs for scenario 3 in~\figref{f:best10VCs5a50} are quite different than scenario 1. Although the cross sections have a flat edge at the top and a sharp vertex at the bottom, they do not resemble a triangle anymore, unlike those in the first two scenarios. Additionally, while the vertical gap sizes are similar, the horizontal gap sizes are greater. \figref{f:optimalVsVCs} shows that the critical viscosity contrast is higher in the optimal design for scenario 3 than scenario 1. This is because the optimal design for scenario 3 induces more cell migration. Although the optimal design for scenario 3 leads the softer cell to migrate more, it does not qualify as one of the optimal designs for scenario 1 since it leads the stiffer cell to displace as well.

In scenario 4, the cells have the same viscosity contrast values as in scenario 1 but the tilt angle is larger. Increasing the tilt angle increases the width of the adjacent stream and leads cells to zig-zag for a wide range of viscosity contrasts~\cite{kabacaoglu-biros17}. That's why, in order for the softer cell to displace cell migration must be stronger than scenario 1. The optimal design for scenario 4 is different than those for the other scenarios. It can induce more migration. When the optimal cross section for scenario 1 is used with the same tilt angle in scenario 4, separation is still possible but the soft cell migrates less.

\paragraph*{Remark} It is also possible to have two cells sorted by using cylindrical pillars and changing the gap sizes only~\cite{zeming-zhang-e16}. For the first scenario, we perform an exhaustive search to find the gap sizes which result in separation using cylindrical pillars with a diameter of $15 \mu m$. We limit the search to the interval $(G_x, G_y) \in [5 \mu m, 15 \mu m]$. Unlike the optimization problems, we did not enforce $\lambda_x = \lambda_y = 25 \mu m$. It turns out that separation is possible for several pairs of gap sizes shown in~\figref{f:exhaustCirc}. However, notice that we do not allow any gap size below $7 \mu m$ in the optimization. The design with equal spacings $G_x = G_y = 7.5 \mu m$ could be chosen by the optimizer but it does not deliver as small objective function value as the optimal designs. That is, it does not provide sorting as efficient as the optimal designs do. The other design with $(G_x, G_y) = (7.5, 10) \mu m$ was not allowed in the optimization problem due to the fixed values of $\lambda_x = 25\mu m$ and $\lambda_y = 25\mu m$.

\subsubsection{HF-DLD simulations}\label{s:fullModelSims}

Recall that in the optimization solve, we used the low-fidelity model. But do the designs work in the high-fidelity model? To answer this question, we performed simulations with HF-DLD. We used a long device which contains $n_{\mathrm{row}} = 12$ rows and $n_{\mathrm{col}} = \left \lceil 4n \right \rceil$ columns where $n$ is the period. For each scenario in~\secref{s:setup} we perform simulations with the initial guess and the optimal designs in Figures~\ref{f:best10VCs5a10}-\ref{f:best10VCs5a10p5}. We present the cell trajectories in Figures~\ref{f:trajsCase1},~\ref{f:trajsCase2},~\ref{f:trajsCase3} and~\ref{f:trajsCase4} for the scenarios 1, 2, 3 and 4, respectively. Here, the cells in blue are softer than the cells in red. We, now, proceed by discussing these results briefly.

\paragraph*{Remark} In order to test the sensitivity to the manufacturing errors, we perturb the B-spline coefficients for the optimal designs by 2\% and perform the simulations again. The cell dynamics are insensitive to this amount of perturbation in all scenarios. We omit these results.

\paragraph*{\bf Scenario 1: $(\nu_1 = 5, \nu_2 = 10)$ and $\theta = 0.17\,\,\mathrm{rad}$ (\figref{f:trajsCase1})}

\begin{figure}[!htb]
\begin{minipage}{\textwidth}
\setcounter{subfigure}{0}
\centering
\renewcommand*{\thesubfigure}{(a)} 
      \hspace{0cm}\subfigure[No-separation with the initial guess]{\scalebox{0.75}{{\includegraphics{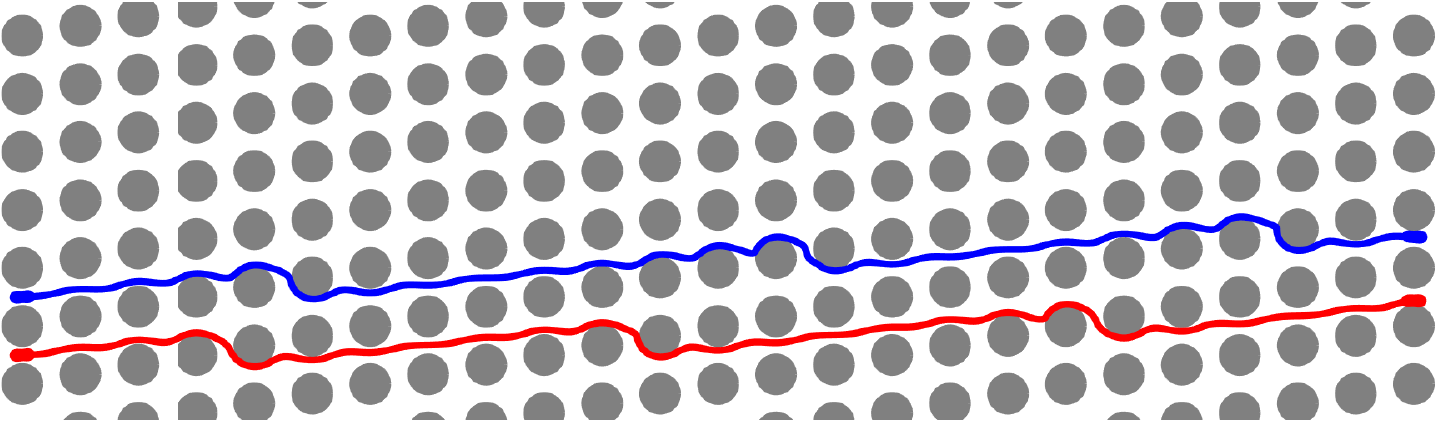}}}
      \label{f:VCs5a10Circ}} 
\end{minipage}
\begin{minipage}{\textwidth}
\setcounter{subfigure}{0}      
\centering
\renewcommand*{\thesubfigure}{(b)} 
      \hspace{0cm}\subfigure[Separation with the optimal design]{\scalebox{0.75}{{\includegraphics{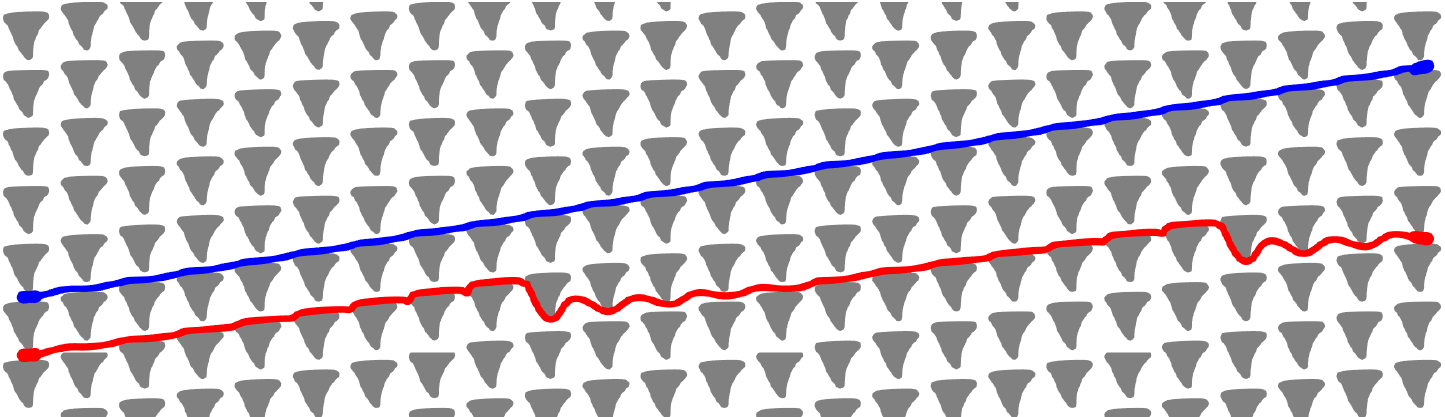}}}
      \label{f:VCs5a10Best}}
\end{minipage}                    
\mcaption{Cell trajectories in HF-DLD for scenario 1: $(\nu_1 = 5, \nu_2 = 10)$ and $\theta = 0.17\,\,\mathrm{rad}$. The cell in blue is softer ($\nu_1 = 5$) than the one in red ($\nu_2 = 10$). The initial guess in~\figref{f:VCs5a10Circ} is in no-separation mode while the optimal design in~\figref{f:VCs5a10Best} leads the soft cell to displace and the stiff cell to zig-zag.}{f:trajsCase1}
\end{figure}
Both cells zig-zag three times in the initial guess (See~\figref{f:VCs5a10Circ}). In the optimal design in~\figref{f:VCs5a10Best} the soft cell always displaces while the stiff cell zig-zags two times. This results in a vertical separation between the cells. \editS{The ratio of the vertical distance between the displacing cell's center and the top of the pillar underneath it at the end of the simulation to the gap is $y_f/G_y = 0.33$.} There are eleven columns between the first and the second zig-zags of the stiff cell (there are less number of columns until it zig-zags for the first time since we initialize cells in the middle of the gap). Therefore, the vertical separation between the cells increases by approximately $\lambda_y = 25 \mu m$ in every eleven columns. The stiff cell zig-zags less frequently in the optimal design than it does in the initial guess because the optimal design increases its vertical migration as well.

\paragraph*{\bf Scenario 2: $(\nu_1 = 4, \nu_2 = 10)$ and $\theta = 0.17\,\,\mathrm{rad}$ (\figref{f:trajsCase2})}

\begin{figure}[!htb]
\begin{minipage}{\textwidth}
\setcounter{subfigure}{0}
\centering
\renewcommand*{\thesubfigure}{(a)} 
      \hspace{0cm}\subfigure[No-separation with the initial guess]{\scalebox{0.75}{{\includegraphics{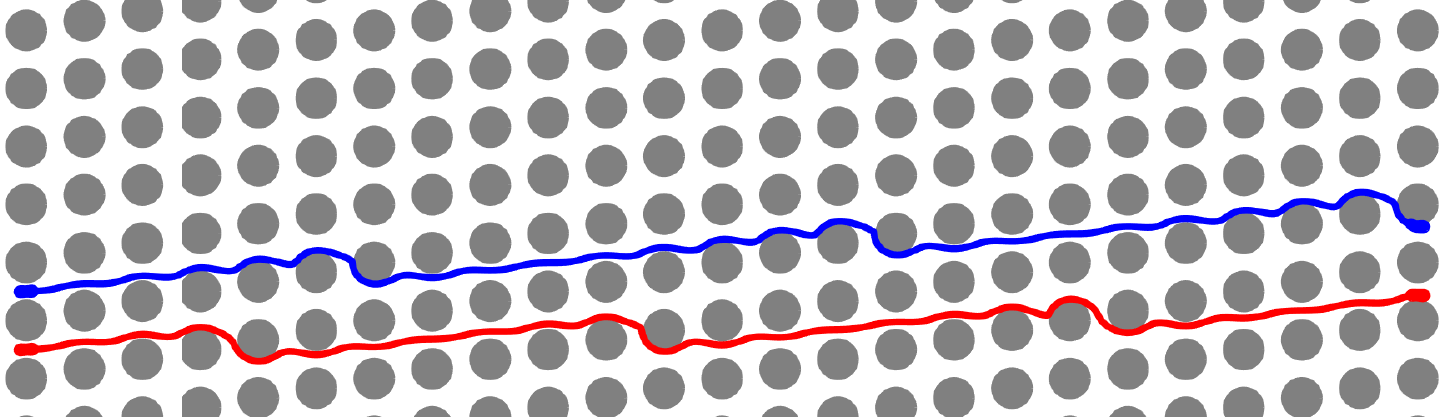}}}
      \label{f:VCs4a10Circ}} 
\end{minipage}
\begin{minipage}{\textwidth}
\setcounter{subfigure}{0}      
\centering
\renewcommand*{\thesubfigure}{(b)} 
      \hspace{0cm}\subfigure[Separation with the optimal design]{\scalebox{0.75}{{\includegraphics{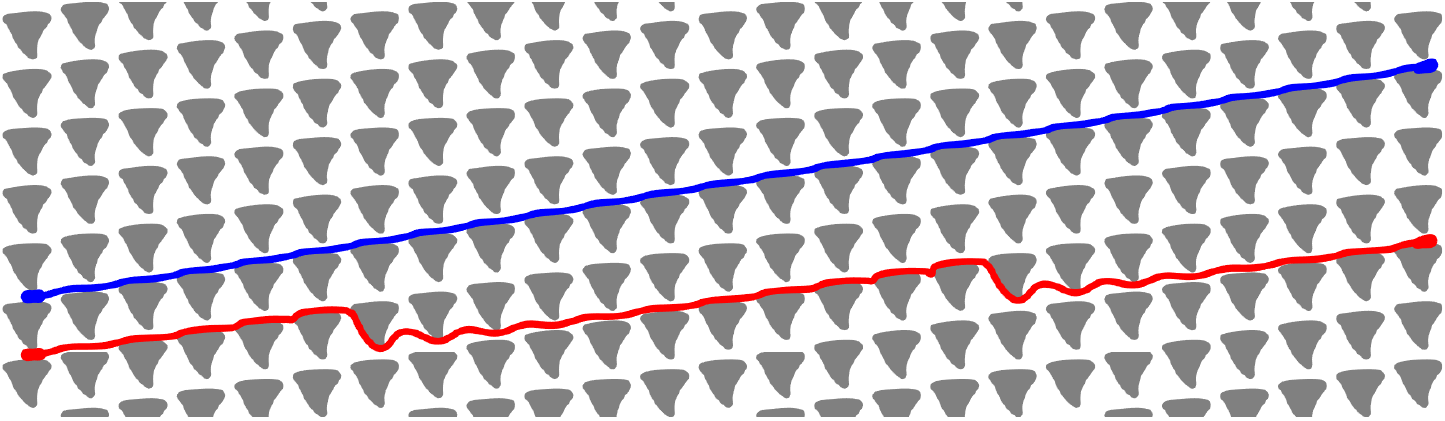}}}
      \label{f:VCs4a10Best}}
\end{minipage}                    
\mcaption{Cell trajectories in HF-DLD for scenario 2: $(\nu_1 = 4, \nu_2 = 10)$ and $\theta = 0.17\,\,\mathrm{rad}$. The cell in blue is softer ($\nu_1 = 4$) than the one in red ($\nu_2 = 10$). The initial guess in~\figref{f:VCs4a10Circ} is in no-separation mode while the optimal design in~\figref{f:VCs4a10Best} leads the soft cell to displace and the stiff cell to zig-zag.}{f:trajsCase2}
\end{figure}
In this scenario, the stiff cell is the same as in the previous scenario. It zig-zags three times in the initial guess (See~\figref{f:trajsCase2}). The soft cell is only a little softer than the previous scenario and also zig-zags three times in the initial guess. In the optimal design in~\figref{f:VCs4a10Best} the soft cell always displaces while the stiff one zig-zags two times. \editS{Here, the ratio of the vertical distance between the soft cell and the pillar to the gap is $y_f/G_y = 0.37$.} The stiff cell zig-zags in every ten columns, which is a little earlier than in the previous scenario. The reason is the following. The soft cell in this scenario can migrate more due to its lower viscosity contrast. Therefore, the optimal design does not need to induce as much migration as in this scenario for the soft cell to displace. This leads the stiff cell to migrate less in the optimal design for this scenario.

\paragraph*{\bf Scenario 3: $(\nu_1 = 5, \nu_2 = 50)$ and $\theta = 0.17\,\,\mathrm{rad}$ (\figref{f:trajsCase3})}

\begin{figure}[!htb]
\begin{center}
\includegraphics[scale=0.75]{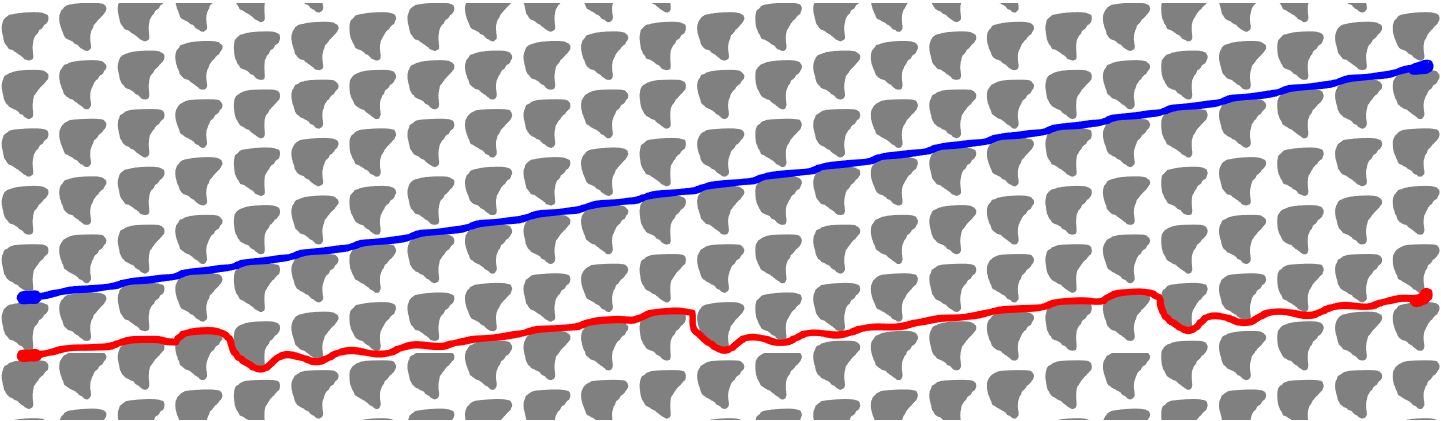} \mcaption{Cell trajectories in HF-DLD with the optimal design for scenario 3: $(\nu_1 = 5, \nu_2 = 50)$ and $\theta = 0.17\,\,\mathrm{rad}$. The cell in blue is softer ($\nu_1 = 5$) than the one in red ($\nu_2 = 50$). The cell trajectories in the initial guess are in~\figref{f:VCs5a50CircMotiv}. The initial guess is in no-separation mode while the optimal design leads the soft cell to displace and the stiff cell to zig-zag.}{f:trajsCase3}
\end{center}
\end{figure}
In this scenario, the soft cell is the same as in the first scenario. The stiff cell has five times greater viscosity contrast value than the previous scenarios. Both cells in this scenario zig-zag three times in the initial guess (See~\figref{f:VCs5a50CircMotiv}). In the optimal design in~\figref{f:trajsCase3} the soft cell displaces while the stiff cell zig-zags three times. \editS{The ratio of the vertical distance between the soft cell and the pillar to the gap is $y_f/G_y = 0.36$.} The stiff cell zig-zags in every seven columns, which is more frequent than the first two scenarios. Although the optimal design for this scenario needs to produce the same vertical migration as in scenario 1 because the soft cells have the same viscosity contrast values, the stiff cell is not much sensitive to this induced migration due to its high viscosity contrast. So, it zig-zags more frequently.

\paragraph*{\bf Scenario 4: $(\nu_1 = 5, \nu_2 = 10)$ and $\theta = 0.2\,\,\mathrm{rad}$ (\figref{f:trajsCase4})}
\begin{figure}[!htb]
\begin{minipage}{\textwidth}
\setcounter{subfigure}{0}
\centering
\renewcommand*{\thesubfigure}{(a)} 
      \hspace{0cm}\subfigure[No separation with the initial guess]{\scalebox{0.75}{{\includegraphics{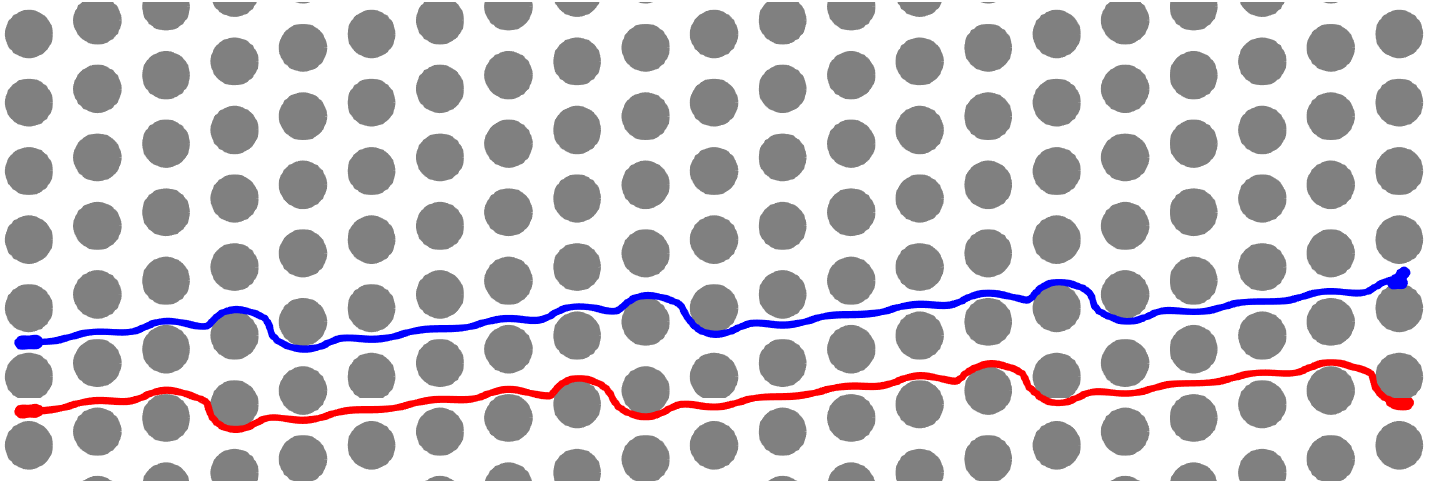}}}
      \label{f:VCs5a10p5Circ}} 
\end{minipage}
\begin{minipage}{\textwidth}
\setcounter{subfigure}{0}      
\centering
\renewcommand*{\thesubfigure}{(b)} 
      \hspace{0cm}\subfigure[Separation with the optimal design]{\scalebox{0.75}{{\includegraphics{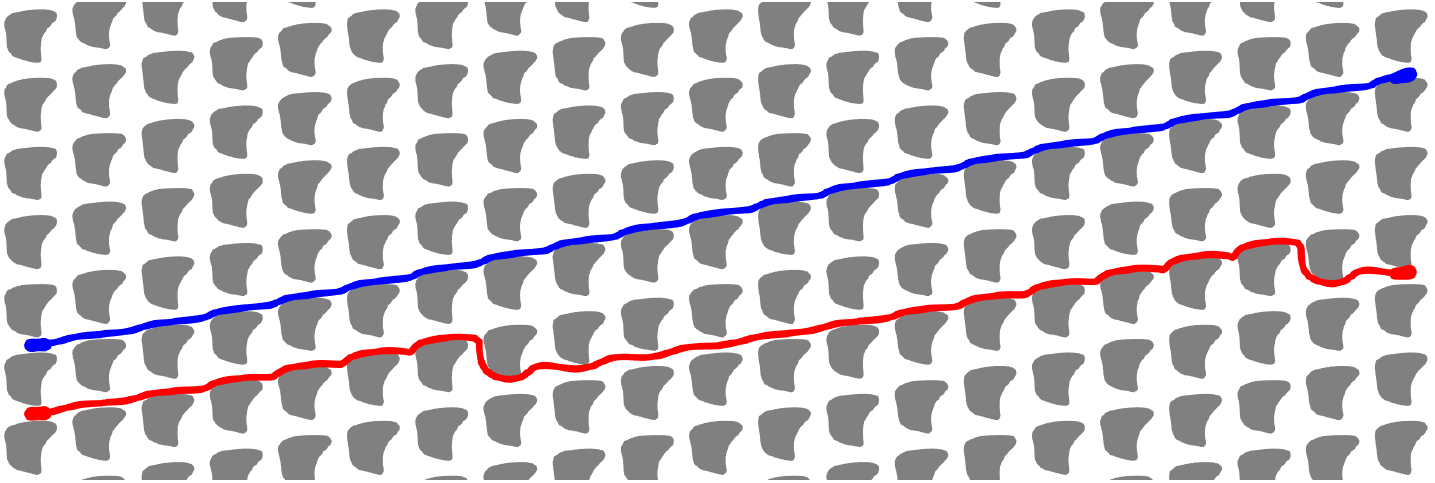}}}
      \label{f:VCs5a10p5Best}}
\end{minipage}                    
\mcaption{Cell trajectories in HF-DLD for scenario 4: $(\nu_1 = 5, \nu_2 = 10)$ and $\theta = 0.2\,\,\mathrm{rad}$. The cell in blue is softer ($\nu_1 = 5$) than the one in red ($\nu_2 = 10$). The initial guess in~\figref{f:VCs5a10p5Circ} does not sort cells while the optimal design in~\figref{f:VCs5a10p5Best} leads the soft cell to displace and the stiff cell to zig-zag.}{f:trajsCase4}
\end{figure}

The only difference between this scenario and the first scenario is the tilt angle of the pillars. The period becomes $n = 5$ in this scenario while it is $n = 6$ for the other scenarios. In the initial guess the soft and the stiff cells zig-zag three and four times (See~\figref{f:VCs5a10p5Circ}). In the optimal design in~\figref{f:VCs5a10p5Best}, the soft cell always displaces and the stiff cell zig-zags in every eleven columns. \editS{The ratio of the vertical distance between the soft cell and the pillar to the gap is $y_f/G_y = 0.30$. This ratio is given to be $0.37$ in the LF-DLD simulation. The LF-DLD and HF-DLD simulations give similar migration of the soft cells in the optimal devices in the other scenarios. The LF-DLD captures the cell dynamics qualitatively in this scenario but the error in the vertical displacement is larger compared to the other scenarios.}  The optimal design induces so much vertical migration that the stiff cell zig-zags less frequently in it compared to the initial guess. This is also observed in scenario 1.

\section{Simplifying pillar cross section\label{s:design}} 

DLD arrays are manufactured by (i) drawing the pillar array pattern using a software, (ii) printing the pattern on a photo mask made of chrome quartz and (iii) manufacturing the arrays based on the mask by an etching technique (e.g., dry or wet etching of a silicon wafer~\cite{mcgrath-bridle-e14}). Printing and etching resolutions determine manufacturing resolution. The printing is done using soft lithography, photolithography or electron beam lithography. These techniques have high resolutions and are capable of printing fine features.  We are not aware of up to what precision the cross sections in the optimal designs shown in Figures ~\ref{f:best10VCs5a10}-\ref{f:best10VCs5a10p5} can be manufactured. The design optimization problem we stated does not have any constraint regarding the manufacturability of the cross sections, which allows the cross sections to have sharp edges and fine features. Even if the cross sections in the optimal designs cannot be manufactured easily, they give an insight to design a device which is (i) easy to manufacture (i.e., with a basic cross section such as square, diamond, triangular without fine features), (ii) robust to uncertainties in the cells' viscosity contrast values and (iii) robust to manufacturing errors. Based on the optimal designs we found and these criteria, we want to design a DLD device with a simpler cross sections for the cells in the first three scenarios. 

\begin{figure}[!htb]
\begin{minipage}{\textwidth}
\setcounter{subfigure}{0}
\centering
\renewcommand*{\thesubfigure}{(a)} 
      \hspace{0cm}\subfigure[Scenario 1: $(\nu_1, \nu_2) = (5, 10)$ and $\theta = 0.17$ rad]{\scalebox{0.75}{{\includegraphics{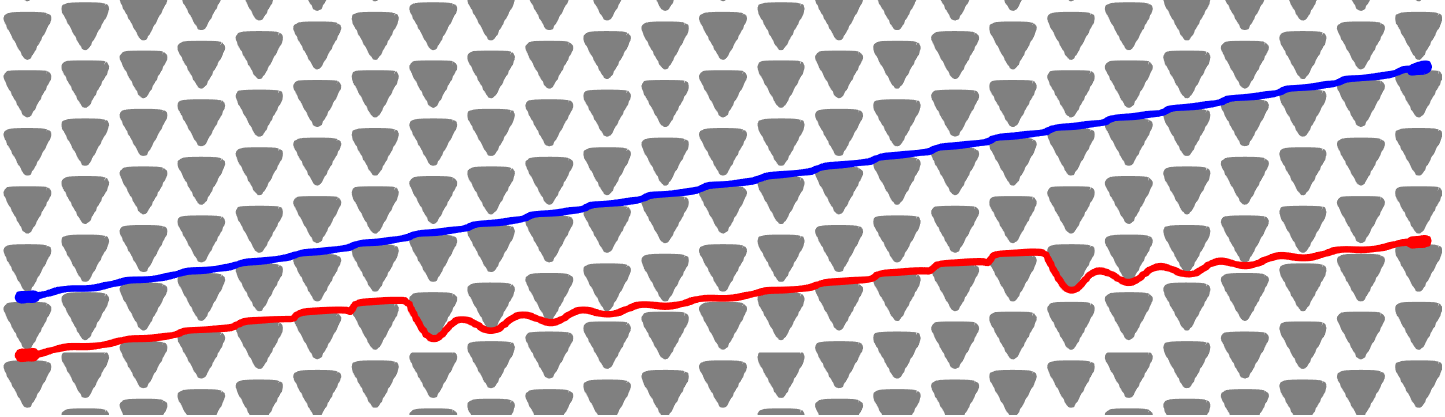}}}
      \label{f:VCs5a10Tri}} 
\end{minipage}
\begin{minipage}{\textwidth}
\setcounter{subfigure}{0}      
\centering
\renewcommand*{\thesubfigure}{(b)} 
      \hspace{0cm}\subfigure[Scenario 2: $(\nu_1, \nu_2) = (4, 10)$ and $\theta = 0.17$ rad]{\scalebox{0.75}{{\includegraphics{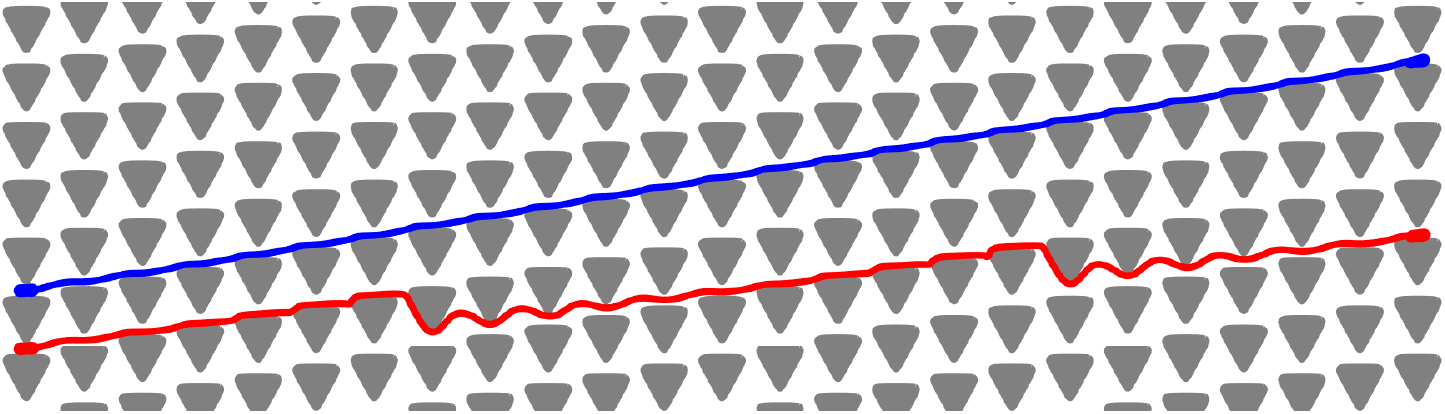}}}
      \label{f:VCs4a10Tri}}
\end{minipage}                    
\mcaption{Cell trajectories in HF-DLD with the proposed triangular pillar cross section for scenarios 1 and 2. See~\figref{f:VCs5a50TriMotiv} for scenario 3. The cell in blue is softer than the one in red. The proposed DLD design is able to sort cells for these scenarios.}{f:trajsForTriPost}
\end{figure}

We set the tilt angle of the pillar rows to $0.17$ rad as in the first three scenarios. The cross sections in the optimal designs for these scenarios have a flat edge at the top and a sharp vertex at the bottom. As discussed in~\secref{s:optimalDesigns}, these features help efficient cell sorting. That's why, we suggest a triangular cross section with such configuration. We need to determine gap sizes and size of the cross section, now. In the optimal designs the horizontal and the vertical gap sizes are in $G_x \in [7, 7.8]\mu m$ and $G_y \in [7, 7.2] \mu m$, respectively. This shows that cell dynamics are not as sensitive to the horizontal gap size as it is to the vertical one. Thus, we first determine the vertical gap size. We want the triangular cross section to have rounded vertices. Rounding the vertex reduces the width of the adjacent stream~\cite{loutherback-sturm-e10}. In order to compensate that, we propose a vertical gap size that is slightly greater than those in the optimal designs. We set the vertical gap size to $G_y = 7.5 \mu m$. To reduce the complexity of the design, We set the horizontal gap size to $G_x = 7.5 \mu m$ as well. Since, we enforce $\lambda_x = \lambda_y = 25 \mu m$, the width and the height of the proposed cross section become $w_{\mathrm{p}} = h_{\mathrm{p}} = 17.5 \mu m$. Based on these properties, we propose a triangular cross section (See~\figref{f:VCs4a10Tri}) of which the coordinates of the B-spline control points are tabulated in~\tabref{t:bSplineCoeffs}.

Let us assess whether the proposed design works for the first three scenarios. We present the cell trajectories in the proposed design in~\figref{f:trajsForTriPost} for scenario 1 and 2, and in~\figref{f:VCs5a50TriMotiv} for scenario 3. In the proposed design, the cells with the viscosity contrasts $\nu = (4,5)$ displace and those with $\nu = (10, 50)$ zig-zag. Both zig-zagging cells zig-zag in every 10 columns. So the cell with $\nu = 10$ zig-zags in the proposed design with the same frequency in the optimal design and the cell with $\nu = 50$ zig-zags less frequently in the proposed design than the optimal design. We compare the objective function values for the proposed design and the optimal designs. Recall that the smaller the value is the more efficient the design is. The objective function is greater for the proposed design than the optimal designs by 22\% for scenario 1, 16\% for scenario 2, 30\% for scenario 3. So although the optimal designs are more efficient, the proposed design has a simpler cross section and is still useful.


\subsection{Sensitivity to uncertainty in viscosity contrast value}\label{s:VCsRobust}

We investigate if the proposed design is sensitive to uncertainties in the cells' viscosity contrast values. We consider the soft and the stiff cells with $(\nu_1, \nu_2) = (5, 10)$, respectively and perturb both $\nu_1$ and $\nu_2$. As mentioned before, it becomes more difficult to sort the cells when their viscosity contrast values are close. For the sensitivity analysis we increase the viscosity contrast value of the soft cell and decrease that of the stiff cell until we find a pair of viscosity contrast values for which the proposed design fails to sort. It turns out that the cells displace for the viscosity contrast $\nu \leq 8$ and zig-zag for $\nu \geq 8.5$. So, the critical viscosity contrast for the separation is between 8 and 8.5 and the closest pair of the viscosity contrast values that the proposed design can separate is $(\nu_1,\nu_2) = (8,8.5)$. In that case, the stiff cell zig-zags with the same frequency as the one with $\nu = 10$ does. Therefore, the proposed design is robust to uncertainties in the cells' viscosity contrast values.

\subsection{Sensitivity to manufacturing errors}\label{s:shapeRobust}

We investigate if the proposed design is sensitive to the manufacturing errors. We consider the manufacturing errors as random perturbations in the coordinates of the B-spline control points of the cross section. While doing that, we still fix $\lambda_x = \lambda_y = 25 \mu m$. That's why, random perturbations in the pillar cross sections result in random perturbations in gap sizes as well. We add a random noise with zero mean and standard deviations of 1\%, 2\%, 5\%, 10\% and 15\% to the coordinates of the B-spline control points in~\tabref{t:bSplineCoeffs}. Then we perform simulations of the cells with $(\nu_1, \nu_2) = (5, 10)$ in the perturbed designs. The proposed design can sort these cells even for 10\% perturbation, which results in 1\% and 5\% changes in the horizontal and vertical gap sizes, respectively. For 15\% perturbation, both cells displace and hence the separation is not possible. Considering the high resolution of the micro manufacturing techniques mentioned before, we conclude that the proposed design is robust to the manufacturing errors.

\section{Conclusion\label{s:conclusions}}
We have posed designing a deterministic lateral displacement device to sort same-size red blood cells by their viscosity contrast values as a design optimization problem. Designing a device amounts to designing a pillar cross section, adjusting center-to-center distances between pillars and tilt angle of the pillar rows. We have parameterized the pillar cross section by uniform $5^{th}$ order B-splines and fix the center-to-center distance and the tilt angle. We have proposed an objective function to try to capture several factors, like device length and sorting efficiency. We have used our 2D model for cell flows through a DLD device, which is based on a boundary integral formulation for the Stokesian particulate flows. We have solved the optimization problem using a stochastic optimization algorithm (CMA-ES). The algorithm converges in $\bigO(1000)$ iterations and each iteration requires the simulation of cell flows in DLD. Additionally, our high-fidelity DLD model is computationally expensive to solve the optimization problem. In order to enable fast solution of the problem, we have proposed a low-fidelity model. We have sought optimal DLD designs for four scenarios which involve cells with close viscosity contrast values. The optimal designs we found can sort the cells and have had different pillar cross sections than the conventional ones (circular, triangular, square or diamond). We have compared the common features of the optimal designs with the designs proposed so far in the literature. While the cells are flowing from left to right, the pillar cross sections in these designs have a flat edge at the top (the shift direction of the pillars) and a sharp vertex at the bottom, which shifts the flow upwards in the vertical gap. The flow shifted upwards induces stronger vertical migration of the cells. However, it also results in a thick adjacent stream, which might cause the cells to zig-zag. The optimal designs compensate that by reducing the vertical gap sizes, which reduces the width of the adjacent stream. Overall, the combination of the pillar cross sections with such orientation and the narrow vertical gaps enables sorting the cells by the viscosity contrast. We have also investigated the sensitivity of the optimal designs to the manufacturing errors and perturbations in the cells' viscosity contrast values. The designs have been robust to the errors and perturbations. Our study demonstrates that solving a design optimization problem systematically discovers optimal DLD designs for high resolution and efficient cell sorting. This is important since otherwise finding a design to sort cells with arbitrary deformability would require an exhaustive search. This study can easily be extended to various design objectives such as reducing the risk of clogging by proposing an appropriate objective function.

\appendix
\section{B-spline coefficients for designs}\label{a:bSplineCoeffs}
We parameterize pillar cross sections using uniform $5^{th}$ order B-splines with 8 control points $c_i$, $i = 1, \cdots, 8$. In Figures~\ref{f:best10VCs5a10}-\ref{f:best10VCs5a10p5}, we present the cross sections in the optimal DLD designs for four scenarios considered in~\secref{s:results}. Additionally, we propose a triangular cross section in~\secref{s:design}. Here, we tabulate the coordinates of the control points for these cross sections in~\tabref{t:bSplineCoeffs} for reproducibility of our results and manufacturing the devices.

\begin{table}[!htb]
\mcaption{Details of the optimal designs for the scenarios considered in~\secref{s:results} and the proposed design in~\secref{s:design}. We tabulate the coordinates of the control points for constructing the pillar cross sections using uniform $5^{th}$ order B-splines $c_i$, $i = 1, \cdots, 8$; the height and the width of the cross sections $h_{\mathrm{p}}$ and $w_{\mathrm{p}}$; the horizontal and vertical gap sizes $G_x$ and $G_y$; the tilt angle of the pillar rows $\theta$. }{t:bSplineCoeffs}
\centering
\begin{tabular}{c | c c | c c | c c | c c | c c}
            & \multicolumn{2}{c|}{Scenario 1} & \multicolumn{2}{c|}{Scenario 2} & \multicolumn{2}{c|}{Scenario 3} & \multicolumn{2}{c}{Scenario 4}& \multicolumn{2}{c}{Proposed} \\
\hline            
Control point & $x (\mu m)$ & $y (\mu m)$ & $x (\mu m)$ & $y (\mu m)$ & $x (\mu m)$ & $y (\mu m)$ & $x (\mu m)$ & $y (\mu m)$ & $x (\mu m)$ & $y (\mu m)$ \\
\hline
$c_1$         & 6.6 & 3.9  & 6.3 & 2.7  & 0.3  & 4.2  & 0.6  & 4.4  & 10.0 & 9.1 \\
\hline
$c_2$         & 15.1 & 6.3 & 11.6 & 6.0 & 13.8 & 11.6  & 14.5  & 10.1 & 0 & 9.1 \\
\hline
$c_3$         & 5.6 & 6.2  & 4.3 & 7.7 &  -2.0 & 10.6  & -1.8  & 10.2 & -10.0 & 9.1 \\
\hline
$c_4$         & -10.6 & 5.6 & -12.8 & 5.1 & -12.0  & 10.4  &  -12.1 & 7.9 & -9.1 & 7.5 \\
\hline
$c_5$         & -2.0 & 4.1  & -6.3 &  4.9 & -6.9 &  -4.3 & -6.9  & 1.4 & -0.9 & -7.5 \\
\hline
$c_6$         & -2.8 & -6.5 & -4.7 &  -6.1 & -6.3 & -0.9  & -10.3 & -6.9 & 0 & -9.1 \\
\hline
$c_7$         & 6.0 & -13.7  & 5.3  & -13.6  & 2.5 & -9.2 & 3.6  & -7.5  & 0.9 & -7.5 \\
\hline
$c_8$         & 1.8 & -10.8  &  1.0 & -8.6  &  1.0 & -6.1 & 2.3  & -9.9 & 9.1 & 7.5 \\
\hline
\hline
$h_{\mathrm{p}} (\mu m)$  & \multicolumn{2}{c|}{18} & \multicolumn{2}{c|}{17.9} & \multicolumn{2}{c|}{17.9} & \multicolumn{2}{c|}{18} & \multicolumn{2}{c}{17.5} \\
\hline
$w_{\mathrm{p}} (\mu m)$  & \multicolumn{2}{c|}{17.5} & \multicolumn{2}{c|}{18} & \multicolumn{2}{c|}{17.3} & \multicolumn{2}{c|}{17.9} & \multicolumn{2}{c}{17.5} \\
\hline
$G_x (\mu m)$   & \multicolumn{2}{c|}{7.5} & \multicolumn{2}{c|}{7.0} & \multicolumn{2}{c|}{7.7} & \multicolumn{2}{c|}{7.1} & \multicolumn{2}{c}{7.5} \\
\hline
$G_y (\mu m)$   & \multicolumn{2}{c|}{7.0} & \multicolumn{2}{c|}{7.1} & \multicolumn{2}{c|}{7.1} & \multicolumn{2}{c|}{7.0} & \multicolumn{2}{c}{7.5} \\
\hline
$\theta$ (rad)  & \multicolumn{2}{c|}{0.17} & \multicolumn{2}{c|}{0.17} & \multicolumn{2}{c|}{0.17} & \multicolumn{2}{c|}{0.2} & \multicolumn{2}{c}{0.17} \\
\end{tabular} 
\end{table}

One can reproduce a cross section in MATLAB using~\algref{a:shapeRepro}. Here, $c$ is a matrix of size $[2\times8]$ and stores the coordinates of 8 B-splines control points in the columns. We also tabulate the size of the cross section ($h_{\mathrm{p}}$ and $w_{\mathrm{p}}$) in~\tabref{t:bSplineCoeffs}, so that, one can scale the produced cross section to match the correct sizes.
\begin{algorithm}[!htb]
\begin{algorithmic} 
\STATE {\% Input the coordinates of the B-spline control points, $c$}
\STATE {$c = [c(:,8);c;c(:,1);c(:,2);c(:,3)]$}
\COMMENT {Repeat the last and the first three control points}
\STATE {$[x,y] = \mathtt{spcrv}(c,5)$}
\COMMENT {Build a $5^{th}$ order spline curve by uniform division}
\STATE {\% $(x,y)$ is a dense sequence of points on the B-spline curve
  representing the cross section.}
\end{algorithmic}
\caption{Reproducing cross sections from B-spline control points}
\label{a:shapeRepro}
\end{algorithm}

\bibliographystyle{plainnat} 
\bibliography{refs}
\biboptions{sort&compress}
\end{document}